Brier-UDAM for Paired Prediction-Rule Comparisons: Latent-Risk Interpretation, Observable Contrasts, and Inference

Eiji Nakatani, Ph.D.[1], Hiroya Hashimoto, Ph.D.[2]

1. Laboratory of Biostatistics, NHO Nagoya Medical Center, Nagoya, Japan.
2. Department of Biostatistics, Nagoya University Graduate School of Medicine, Nagoya, Japan.

**Corresponding author:**
Eiji Nakatani, Ph.D.
Laboratory of Biostatistics, Clinical Research Center,
NHO Nagoya Medical Center, Nagoya, Japan.
Email: nakatani.eiji.int@gmail.com

**Abstract**

The Brier risk summarizes probability-prediction error but does not explain why two prediction rules differ. We develop Brier-UDAM, a framework for paired binary prediction-rule comparisons that separates the Brier-risk difference into mean-bias, dispersion, and Brier-scale alignment contributions. Although the observable contrast identity follows algebraically from Yates-type moment decompositions, the framework gives these terms a latent-risk interpretation and shows that paired contrasts can be identified from observed outcomes and paired predictions without estimating the conditional event-risk function. For two rules evaluated in the same target population, the exact identity $\Delta R = \Delta M + \Delta D + \Delta L$ holds, and the corresponding plug-in estimators reproduce the empirical paired Brier-risk difference exactly in finite samples. We derive influence-function-based inference for first-order regular contrasts and a projection-based confidence procedure for the mean-bias contrast when first-order inference degenerates. Simulations confirmed interpretable component behavior, recovery of the target contrasts, and generally satisfactory inference in regular settings; projection intervals retained coverage near degeneracy but were conservative. In a large public-data analysis of 30-day hospital readmission, models with similar held-out Brier risks showed substantially larger and opposing dispersion and alignment contributions, revealing cancellation that was hidden in the aggregate score. Brier-UDAM therefore provides a diagnostic account of paired Brier-risk differences, while calibration, discrimination, and information-based assessments continue to address their distinct performance questions.

## 1. Introduction

Clinical prediction models increasingly provide individualized event probabilities to support risk communication, clinical monitoring, and treatment decisions. Their performance cannot be characterized by discrimination alone. A model may correctly rank individuals from lower to higher risk while systematically reporting probabilities that are too high, too low, or insufficiently variable. Calibration and discrimination therefore describe distinct aspects of predictive performance and should be evaluated separately when assessing probabilistic predictions.[1–3]

The Brier score is widely used as an overall measure of probabilistic prediction performance for binary outcomes.[4] It is defined as the mean squared difference between the predicted probability and the observed outcome.4 As a strictly proper scoring rule, its population expectation is uniquely minimized by the true conditional event probability.[5] The Brier score therefore provides a coherent summary of probability-prediction error that reflects aspects of both calibration and discrimination. It is not, however, a direct measure of clinical utility, which requires explicit consideration of the consequences of alternative clinical decisions.[6]

Despite these desirable properties, the Brier score is an aggregate measure. A difference in Brier risk indicates that two prediction rules differ in overall prediction error but does not explain why. The difference may arise from their average predicted risks, the spread of their predicted probabilities, or the way prediction errors vary across individuals with different underlying risks. These mechanisms have different implications for interpretation, recalibration, and model revision, yet may produce similar total Brier risks.

Several decompositions have been developed to improve the interpretability of the Brier score. Murphy's decomposition organizes the score into uncertainty, reliability, and resolution by conditioning on issued predictions,[7,8] and subsequent work has refined this framework through grouping effects, within-group variation, reliability, and conditional information.[9–12] Other approaches describe forecast performance through the means, variances, covariances, or conditional distributions of observed outcomes and predictions.[13–15] In particular, Yates-type decompositions express the Brier score through outcome variance, prediction variance, prediction–outcome covariance, and squared mean error.[13,15] A third perspective concerns the information preserved or lost when richer predictor information is compressed into an issued probability. These approaches describe different aspects of prediction performance: prediction-conditioned decompositions emphasize calibration and realized resolution, moment-based decompositions characterize score differences through observed moments and associations, and information-based decompositions concern preservation of risk information.

When these decompositions are used to interpret differences between prediction rules, a specific issue arises. Quantities describing dispersion relative to the true conditional event risk or the alignment of prediction error with that risk are naturally interpretable at the model level, but the individual-level conditional risk is unobserved. Such model-specific quantities cannot in general be recovered from outcomes and issued predictions alone. At the same time, Yates-type moment identities suggest that some between-model quantities may be obtainable without estimating the conditional-risk function. This creates a distinction between model-level latent-risk interpretation and observed-data identification in paired prediction-rule comparisons.

For two prediction rules evaluated in the same target population, common latent-risk terms cancel and the remaining between-model quantities admit exact observed-data representations. Paired differences in mean bias, prediction dispersion, and risk-error alignment can therefore be related directly to the observed Brier-risk difference without estimating the conditional event risk or introducing an additional reference model.

We develop Brier-UDAM as a latent-risk interpretation and identification-and-inference framework for paired Brier-risk contrasts. The framework distinguishes mean bias, dispersion mismatch, and risk-error alignment; establishes the identification status of model-specific quantities and paired contrasts; provides finite-sample estimators whose contributions exactly reproduce the paired empirical Brier-risk difference; and derives influence-function-based inference for first-order regular contrasts. We further characterize first-order degeneracy of the mean-bias contrast and develop a projection-based confidence procedure for that setting. Simulation studies assess component interpretation, finite-sample recovery, and inferential operating characteristics, and an empirical analysis illustrates the practical interpretation of the paired decomposition.

## 2. Methods

### 2.1 Setup, target population, and notation

Let $O = (Y,X)$ denote an observation drawn from a specified target population, where $Y \in \{0,1\}$ is a binary outcome and $X$ is a vector of predictors. All expectations, variances, and covariances in this study are defined with respect to the joint distribution of $(Y, X)$ in this target population.

The true conditional event probability is defined as

$$r(X) = P(Y = 1|X) = E(Y|X).$$

The function $r(X)$ represents the individual-level event risk determined by the full predictor information $X$. It is a population-level function and is not directly observed for an individual subject.

Let $p(X) \in [0,1]$ denote a probabilistic prediction rule. The prediction rule may be obtained from a parametric, semiparametric, or machine-learning model. No assumption is made that $p(X)$ is correctly specified or belongs to the same functional class as $r(X)$.

The population Brier risk of $p$is defined as

$$R(p) = E\left[\{Y - p(X)\}^2\right].$$

This is the expected squared difference between the observed binary outcome and the predicted probability in the target population. Because the Brier score is a strictly proper scoring rule, $R(p)$is minimized by the true conditional risk $p(X) = r(X)$.

For later decomposition, define the prediction error relative to the conditional event risk as $e_p(X) = p(X) - r(X)$.

The subscript $p$emphasizes that the error depends on the prediction rule being evaluated. When only one prediction rule is under consideration, we write $e(X)$ for simplicity.

For comparison of two prediction rules, let $p_1(X)$, $p_2(X) \in [0,1]$ denote two probabilistic predictions defined for the same individuals in the same target population. Their population Brier-risk contrast is defined as

$$\Delta R_{12} = R(p_1) - R(p_2).$$

Throughout this study, contrasts are defined as the first prediction rule minus the second. Therefore, $\Delta R_{12} > 0$ indicates that $p_1$has a larger Brier risk, and hence poorer overall probability prediction performance, than $p_2$ in the target population.

The estimands considered in this study describe the performance of prespecified prediction rules in a common target population. Consequently, they depend not only on the prediction functions themselves but also on the target-population distribution of $X$. The same pair of prediction rules may therefore have different Brier risks and different decomposition components in populations with different outcome prevalence, predictor distributions, or case mix.

### 2.2 Separation of irreducible uncertainty and structural prediction error

To separate outcome randomness from model-dependent prediction error, write

$$Y - p(X) = \{Y - r(X)\} - \{p(X) - r(X)\}.$$

Squaring both sides gives

$$\{Y - p(X)\}^2 = \{Y - r(X)\}^2 + \{p(X) - r(X)\}^2 - 2\{Y - r(X)\}\{p(X) - r(X)\}.$$

Because $p(X) - r(X)$is a function of $X$,

$$E\big[\{Y - r(X)\}\{p(X) - r(X)\}\big] = E\big[\{p(X) - r(X)\}E\{Y - r(X)|X\}\big] = 0,$$

where $E\{Y - r(X)|X\} = 0$. Therefore,

$$R(p) = E\big[\{Y - r(X)\}^2\big] + E\big[\{p(X) - r(X)\}^2\big].$$

Since $Y|X$ follows a Bernoulli distribution with probability $r(X)$,

$$E\big[\{Y - r(X)\}^2|X\big] = \mathrm{Var}(Y|X) = r(X)\{1 - r(X)\}.$$

It follows that

$$R(p) = U + T(p), \qquad (1)$$

where $U = E\big[r(X)\{1 - r(X)\}\big]$ is the irreducible uncertainty and $T(p) = E\big[\{p(X) - r(X)\}^2\big]$ is the structural prediction error.

The irreducible uncertainty represents random Bernoulli variation among individuals with the same predictor values. It does not depend on the prediction rule. In contrast, $T(p)$ measures the squared distance between the prediction rule and the conditional event risk function and excludes realized outcome noise.

### 2.3 The Brier-UDAM decomposition

Define the prediction error relative to the conditional event risk as $e(X) = p(X) - r(X)$. Then $T(p) = E\{e(X)^2\}$.

Using the mean–variance identity,

$$T(p) = [E\{e(X)\}]^2 + \mathrm{Var}\,\{e(X)\}. \qquad (2)$$

The first term is

$$[E\{e(X)\}]^2 = [E\{p(X) - r(X)\}]^2.$$

For the variance term,

$$\text{Var}\,\{e(X)\} = \text{Var}\,\{p(X) - r(X)\} = \text{Var}\,\{p(X)\} + \text{Var}\,\{r(X)\} - 2\text{Cov}\,\{p(X), r(X)\}.$$

To express this variance in terms of a dispersion difference and an error–risk association, note that

$$\text{Cov}\left\{r(X), p(X) - r(X)\right\} = \text{Cov}\left\{r(X), p(X)\right\} - \text{Var}\left\{r(X)\right\}.$$

Therefore,

$$\text{Var}\,\{e(X)\} = \text{Var}\,\{p(X)\} - \text{Var}\,\{r(X)\} - 2\text{Cov}\,[r(X), p(X) - r(X)] \tag{3}$$

Combining equations (2) and (3) gives

$$T(p) = M(p) + D(p) + L(p) = M(p) + D(p) - 2A(p). \tag{4}$$

Here $M(p) = \left[E\left\{p(X) - r(X)\right\}\right]^2$ is squared mean bias, $D(p) = Var\left\{p(X)\right\} - Var\left\{r(X)\right\}$ is dispersion mismatch, $A(p) = Cov\left[r(X), p(X) - r(X)\right]$ is the alignment term, and $L(p) = -2A(p)$ is its signed contribution on the Brier-risk scale.

Consequently,

$$R(p) = U + D(p) - 2A(p) + M(p) = U + M(p) + D(p) + L(p). \tag{5}$$

Equation (5) defines the Brier-UDAM decomposition: uncertainty (U), dispersion (D), alignment (A), and mean bias (M). For empirical accounting we report $L = -2A$, because $M$, D, and $L$ then sum directly to the model-dependent Brier risk and to paired Brier-risk contrasts. $L$ is a reporting convention, not a fifth UDAM term.

### 2.4 Interpretation and sign structure of the UDAM terms

**Mean bias (M) and signed mean error**

The mean-bias component is $M(p) = \left[E\left\{p(X) - r(X)\right\}\right]^2$. Because $E\left\{r(X)\right\} = E(Y)$, it can equivalently be written as $M(p) = \left[E\left\{p(X)\right\} - E(Y)\right]^2$. We define the signed mean error as $m(p) = E\left\{p(X) - r(X)\right\} = E\left\{p(X)\right\} - E(Y)$, so that $M(p) = \left\{m(p)\right\}^2$. Equivalently, $m(p) = \text{sign}\left[E\left\{p(X)\right\} - E(Y)\right]\sqrt{M(p)}$. Thus, $M(p)$ quantifies the magnitude of the discrepancy between the mean predicted probability and the population event probability, whereas $m(p)$retains its direction. A positive signed mean error indicates average overprediction, whereas a negative value indicates average underprediction.

For two prediction rules $p_1$and $p_2$, let $m_j = m(p_j) = E\{p_j(X)\} - E(Y), j = 1,2$. The mean-bias contrast is then $\Delta M_{12} = m_1^2 - m_2^2$. The corresponding signed mean-error contrast is $\Delta m_{12} = m_1 - m_2 = E\{p_1(X) - p_2(X)\}$. The direction of the model-specific mean errors cannot be recovered from $\Delta M_{12}$ alone. Therefore, $m_1$and $m_2$, or equivalently $\Delta m_{12}$together with one of the model-specific signed errors, should be reported when the direction of systematic overprediction or underprediction is relevant. A signed square root of $\Delta M_{12}$ should not be used because $\Delta M_{12}$ is the difference between two squared errors and is not itself the square of a single signed quantity.

**Dispersion mismatch**

The dispersion component is $D(p) = \mathrm{Var}\{p(X)\} - \mathrm{Var}\{r(X)\}$. It compares the dispersion of predicted probabilities with the dispersion of the latent conditional risks. A positive value indicates that the predicted probabilities are more dispersed than the latent risks, whereas a negative value indicates that they are less dispersed. This component is a signed variance difference, not a non-negative loss.

**Alignment (A) and its Brier-scale contribution (L=-2A)**

The risk-alignment contribution is

$$A(p) = Cov\{r(X), p(X) - r(X)\}, \qquad L(p) = -2A(p).$$

The alignment term describes how prediction error varies across the conditional-risk distribution. If $p(X) - r(X)$ tends to increase with $r(X)$, $A(p)$ is positive and its Brier-scale contribution $L(p)$ is negative; if the error tends to decrease with $r(X)$, $A(p)$ is negative and $L(p)$ is positive.

Because alignment is structurally coupled with dispersion, neither $A(p)$ nor $L(p)$ should be interpreted in isolation as indicating better or worse predictive performance.

Although

$$M(p) \geq 0, \qquad M(p) + D(p) - 2A(p) = M(p) + D(p) + L(p) = T(p) \geq 0,$$

but $D(p)$, $A(p)$, and $L(p)$ may be negative. Brier-UDAM is therefore a signed accounting decomposition rather than a decomposition into separate non-negative losses.

**2.5 Structural evolution under nested information sets**

Under progressively richer information sets, the Brier-UDAM decomposition has a simple interpretation. This setting formalizes the sequential addition of prognostic factors to a prediction model.

Let

$$\mathcal{F}_0 \subseteq \mathcal{F}_1 \subseteq \cdots \subseteq \mathcal{F}_K \subseteq \sigma(X)$$

denote nested information sets, and define the optimal prediction based on $\mathcal{F}_k$ as

$$p_k = E(Y|\mathcal{F}_k) = E\{r(X)|\mathcal{F}_k\}.$$

The initial information set $\mathcal{F}_0$may be taken to be trivial, in which case

$$p_0 = E(Y)$$

is the null prediction assigning the marginal event probability to every individual. Because conditional expectation preserves the marginal mean,

$$E(p_k) = E\{r(X)\},$$

and therefore the mean-bias component is zero at every stage:

$$M_k = 0.$$

Moreover, the projection property of conditional expectation gives

$$\mathrm{Cov}\{r(X), p_k\} = \mathrm{Var}(p_k).$$

It follows that

$$D_k = \mathrm{Var}(p_k) - \mathrm{Var}\{r(X)\},$$

$$A_k = D_k,$$

and hence

$$L_k = -2A_k = -2D_k$$

The structural prediction error is therefore

$$T_k = M_k + D_k + L_k = \mathrm{Var}\{r(X)\} - \mathrm{Var}(p_k).$$

This identity clarifies that an increase in prediction dispersion does not necessarily indicate deterioration. Under ideal prediction, additional prognostic information allows predicted risks to vary more appropriately across individuals and thereby corrects underdispersion relative to the true conditional-risk distribution.

For two successive information sets, define

$$\Delta V_k = \mathrm{Var}(p_k) - \mathrm{Var}(p_{k-1}).$$

Since conditional expectations under nested information sets are successive $L_2$ projections,

$$\Delta V_k \geq 0.$$

Consequently,

$$\Delta M_k = 0, \qquad \Delta D_k = \Delta V_k, \qquad \Delta A_k = \Delta D_k, \qquad \Delta L_k = -2\Delta V_k,$$

and

$$\Delta T_k = -\, \Delta V_k = -\, \Delta D_k = \tfrac{1}{2}\Delta L_k.$$

Under optimal prediction based on nested information sets, adding prognostic information increases both D and A by the recovered risk variation, while the reported alignment contribution L decreases by twice that amount. The resulting reduction in Brier risk equals the increase in prediction dispersion. This is an ideal projection benchmark, not a general property of empirically fitted or unrelated models.

Empirically fitted models need not satisfy this projection property. For a general sequence of fitted prediction models,

$$\Delta T_k = \Delta M_k + \Delta D_k + \Delta L_k,$$

and the ideal relation $\Delta L_k = -2\Delta D_k$ provides a benchmark for comparison with general fitted-model sequences. An increase in dispersion is favorable only when accompanied by sufficient improvement in alignment and not offset by deterioration in mean bias. Departures from the ideal relation help distinguish recovery of genuine risk heterogeneity from prediction spreading without corresponding risk alignment. Formal results are provided in Appendix C.

### 2.6 Alignment–dispersion relationship along a linear prediction path

The nested-information result describes an ideal sequence of conditional-risk predictions. To examine the corresponding relationship for two arbitrary prediction functions, consider the linear path

$$p_w(X) = p_0(X) + w\{p_1(X) - p_0(X)\}, \qquad 0 \le w \le 1,$$

This path is used only as an analytical device and does not define a model-combination procedure.

Along this path, the alignment contribution changes linearly with w, whereas the dispersion contribution changes quadratically. Movement toward one prediction rule may improve risk alignment while simultaneously increasing prediction dispersion. At an interior Brier-optimal point, the marginal improvement in alignment is balanced by the combined marginal changes in mean bias and dispersion; when the two prediction functions have the same mean predicted risk, the balance is between alignment and dispersion alone.

This relationship also explains why two prediction rules with similar overall Brier risks may nevertheless have substantially different dispersion and alignment contributions. Formal derivations are provided in Appendix D.

### 2.7 Relation to established Brier-score decompositions

Murphy's decomposition organizes the Brier risk in terms of uncertainty, reliability, and resolution by conditioning on the issued prediction. In population notation, with $q(p) = E(Y|p)$, a common form is

$$R(p) = E\left[\{p - q(p)\}^2\right] - \mathrm{Var}\left\{q(p)\right\} + \mathrm{Var}(Y) .$$

This formulation evaluates agreement between predictions and observed event frequencies among individuals receiving similar predicted probabilities.

A Yates-type moment decomposition instead expresses the Brier risk through observed moments:

$$R(p) = \mathrm{Var}(Y) + \mathrm{Var}\left\{p(X)\right\} - 2\,\mathrm{Cov}\left\{Y,p(X)\right\} + \left[E(Y) - E\{p(X)\}\right]^2.$$

Subtracting this expression for $p_2$ from that for $p_1$ eliminates the common outcome-variance term and gives

$$\begin{aligned} R(p_1) - R(p_2) = &\left[\left\{E(p_1) - E(Y)\right\}^2 - \left\{E(p_2) - E(Y)\right\}^2\right] \\ &+ \left[Var(p_1) - Var(p_2)\right] \\ &-2Cov\{Y,p_1 - p_2\}. \end{aligned}$$

The observable contrast identity is algebraically equivalent to the difference between two Yates-type moment decompositions. The uncensored-score expansion itself is established. Brier-UDAM connects that moment identity to a latent-risk parameterization, distinguishes model-specific quantities that depend on the conditional-risk function from paired contrasts having reduced-data representations, establishes an exact finite-sample accounting identity for those paired contrasts, and provides an inferential framework with influence-function-based inference in first-order regular settings and projection-based inference for the first-order-degenerate mean-bias contrast.

Murphy-type, moment-based, and information-set decompositions answer different questions. They address prediction-conditioned calibration and resolution, observed moment structure, and information preservation, respectively. Brier-UDAM addresses a narrower question: how a paired Brier-risk difference is assembled from mean-bias, prediction-dispersion, and Brier-scale alignment contributions.

### 2.8 Model-specific dependence on the conditional risk and reduced-data identification of paired contrasts

The model-specific Brier-UDAM quantities are defined relative to the conditional event risk $r(X)$. The mean-bias component

$$M(p) = [E\{p(X) - r(X)\}]^2$$

is identified from the observed-data distribution because $E\{r(X)\} = E(Y)$, so that

$$M(p) = [E\{p(X)\} - E(Y)]^2.$$

In contrast, the model-specific dispersion and alignment quantities depend on the conditional-risk function $r(X)$ and are not identified from the reduced distribution of $\{Y, p(X)\}$ alone. When X is observed, these quantities are functionals of the full observed-data law through $r(X) = E(Y|X)$, but their estimation requires estimation of that conditional-risk function. The paired contrasts developed below admit reduced-data representations in terms of $\{Y, p_1(X), p_2(X)\}$ and therefore avoid this nuisance-function estimation.

Now let $p_1(X)$ and $p_2(X)$ be two prediction rules evaluated for the same individuals in the same target population. Define

$$\Delta R_{12} = R(p_1) - R(p_2),$$

and the model-dependent contribution contrasts

$$\Delta M_{12} = M(p_1) - M(p_2),$$
$$\Delta D_{12} = D(p_1) - D(p_2),$$

and

$$\Delta L_{12} = L(p_1) - L(p_2).$$

Throughout, contrasts are defined as the first prediction rule minus the second, so that $\Delta R_{12} > 0$ indicates a larger population Brier risk for $p_1$. Because the irreducible uncertainty $U$ is common to both prediction rules,

$$\Delta R_{12} = \Delta M_{12} + \Delta D_{12} + \Delta L_{12}. \qquad (6)$$

**Proposition 1 (Identification of paired Brier-UDAM contrasts).**

Under the setup of Section 2.1, let$p_1(X)$ and $p_2(X)$ be two probabilistic prediction rules evaluated for the same individuals in the same target population. Then the paired contrasts $\Delta M_{12}$, $\Delta D_{12}$, and $\Delta L_{12}$, and hence $\Delta R_{12}$, are identified from the observed-data distribution of $\{Y, p_1(X), p_2(X)\}$, even though their model-specific forms depend on $r(X)$ and are not identified from the reduced prediction–outcome distribution alone. Specifically,

$$\Delta M_{12} = [E\{p_1(X)\} - E(Y)]^2 - [E\{p_2(X)\} - E(Y)]^2, \quad (7)$$

$$\Delta D_{12} = \text{Var}\{p_1(X)\} - \text{Var}\{p_2(X)\}, \quad (8)$$

and

$$\Delta L_{12} = -2\text{Cov}\{Y, p_1(X) - p_2(X)\}. \quad (9)$$

**Proof.**

To establish (7), note that $E\{r(X)\} = E(Y)$. For (8), the common term $\text{Var}\{r(X)\}$ cancels between the two model-specific dispersion quantities. For the alignment contribution,

$$\Delta A_{12} = \text{Cov}\{r(X), p_1(X) - p_2(X)\}.$$

For any integrable function $h(X)$,

$$\text{Cov}\{Y, h(X)\} = \text{Cov}\{r(X), h(X)\},$$

because

$$E[Yh(X)] = E[h(X)E(Y|X)] = E\{h(X)r(X)\},$$

and $E(Y) = Er(X)$. Taking $h(X) = p_1(X) - p_2(X)$ yields

$$\Delta A_{12} = Cov\{Y, p_1(X) - p_2(X)\},$$

and therefore equation

$$\Delta L_{12} = -2Cov\{Y, p_1(X) - p_2(X)\},$$

which establishes equation (9). The covariance identity used in this derivation is stated formally and proved in Appendix A. This completes the proof.

Proposition 1 establishes the central distinction underlying Brier-UDAM: model-specific dispersion and alignment quantities are defined relative to the conditional-risk function and are not identified from the reduced prediction-outcome distribution alone, whereas their paired between-model contrasts admit exact reduced-data representations without estimating that function. The primary reporting estimands are therefore $\Delta R_{12}$, $\Delta M_{12}$, $\Delta D_{12}$, and $\Delta L_{12} = -2\Delta A_{12}$. Figure 1 summarizes this distinction.

## 2.9 Sample estimators and computation

Suppose an evaluation sample contains $\{Y_i, p_{1i}, p_{2i}\}_{i=1}^{n}$, where $p_{1i}$ and $p_{2i}$ are predictions from two prespecified prediction rules for the same subject $i$. The paired structure is required because the proposed contrasts are defined for two prediction rules in the same target population and are estimated using the subject-level prediction difference $p_{1i} - p_{2i}$ for the same evaluation subjects.

Let

$$\bar{Y} = \frac{1}{n}\sum_{i=1}^{n} Y_i \,, \; \bar{p}_j = \frac{1}{n}\sum_{i=1}^{n} p_{ji} \,, j = 1,2.$$

For the exact finite-sample decomposition, define empirical variances and covariances using denominator $n$:

$$\mathrm{Var}_{\,n}(Z) = \frac{1}{n}\sum_{i=1}^{n} (Z_i - \bar{Z})^2,$$

and

$$\mathrm{Cov}_{\,n}(Z,W) = \frac{1}{n}\sum_{i=1}^{n} (Z_i - \bar{Z})(W_i - \bar{W}).$$

The plug-in estimator of the mean-bias contrast is

$$\widehat{\Delta M}_{12} = (\bar{p}_1 - \bar{Y})^2 - (\bar{p}_2 - \bar{Y})^2. \qquad (10)$$

The plug-in estimator of the dispersion contrast is

$$\widehat{\Delta D}_{12} = \mathrm{Var}_{\,n}(p_1) - \mathrm{Var}_{\,n}(p_2). \qquad (11)$$

The plug-in estimator of the Brier-scale alignment contribution is

$$\widehat{\Delta L}_{12} = -2\mathrm{Cov}_{\,n}(Y,p_1 - p_2). \qquad (12)$$

The empirical total Brier-risk contrast is

$$\widehat{\Delta R}_{12} = \frac{1}{n}\sum_{i=1}^{n} [(Y_i - p_{1i})^2 - (Y_i - p_{2i})^2]\,. \qquad (13)$$

With the same denominator-$n$ convention used throughout,

$$\widehat{\Delta R}_{12} = \widehat{\Delta M}_{12} + \widehat{\Delta D}_{12} + \widehat{\Delta L}_{12} \qquad (14)$$

holds exactly in every finite sample, apart from numerical rounding. A direct algebraic verification of this finite-sample identity is provided in Appendix B.

For prediction rule $p_j$, the signed mean error is estimated by $m_j = mean(p_j) - mean(Y)$, $j = 1,2$. The model-specific empirical mean-bias quantity is $M_j = {m_j}^2$, and the mean-bias contrast in equation (10) is $\Delta M_{12} = {m_1}^2 - {m_2}^2$. The signed mean-error contrast is $\Delta m_{12} = m_1 - m_2 = mean(p_1) - mean(p_2)$. Thus, $\Delta m_{12}$ does not depend on the observed event proportion because the common outcome mean cancels. The model-specific $m_1$ and $m_2$ should still be reported when the direction of overprediction or underprediction is relevant.

Computation of the paired Brier-UDAM contrasts requires only the sample means of the outcome and predictions, the sample variances of the two prediction sets, and the sample covariance between the outcome and the subject-level prediction difference $p_1 - p_2$.

The calculations must be based on predictions generated for the same evaluation subjects so that the comparison remains paired.

For the evaluation of predictive performance, predictions should preferably be generated out of sample, for example in an independent test set or through an appropriate cross-validation or cross-fitting procedure. This validation requirement is distinct from the algebraic decomposition: the decomposition can be calculated for any paired predictions, whereas in-sample predictions may provide optimistically biased estimates of performance in new individuals.

**2.10 Sampling uncertainty and large-sample inference**

The component estimators are smooth functions of empirical first and second moments and are consistent under independent and identically distributed sampling. In first-order regular settings, they admit asymptotically linear representations with the influence functions given below.

Define

$$\mu_Y = E(Y),\ \mu_j = E\{p_j(X)\}, \qquad j = 1,2,$$

and

$$d(X) = p_1(X) - p_2(X),\ \mu_d = E\{d(X)\}.$$

The influence function for the mean-bias contrast is

$$\begin{aligned} \phi_M(O) &= 2(\mu_1 - \mu_Y)[p_1 - \mu_1 - (Y - \mu_Y)] \\ &\quad -2(\mu_2 - \mu_Y)[p_2 - \mu_2 - (Y - \mu_Y)]. \end{aligned} \tag{15}$$

The influence function for the dispersion contrast is

$$\phi_D(O) = [(p_1 - \mu_1)^2 - \mathrm{Var}\,(p_1)] - [(p_2 - \mu_2)^2 - \mathrm{Var}\,(p_2)]. \tag{16}$$

Because

$$\Delta L_{12} = -2\{E(Yd) - \mu_Y \mu_d\},$$

the influence function for the Brier-scale alignment contribution is

$$\varphi_L(O) = -2\big[(Y - \mu_Y)(d - \mu_d) - Cov(Y,d)\big]. \qquad (17)$$

The influence function for the total Brier-risk contrast is

$$\phi_R(O) = \phi_M(O) + \phi_D(O) + \phi_L(O). \qquad (18)$$

Equivalently,

$$\phi_R(O) = [(Y - p_1)^2 - (Y - p_2)^2] - \Delta R_{12}. \qquad (19)$$

The influence function for the model-specific signed mean error $m_j = E\{p_j(X) - Y\}$ is

$$\phi_{m_j}(O) = p_j(X) - Y - m_j, j = 1,2.$$

The influence function for the signed mean-error contrast is

$$\phi_{\Delta m}(O) = \phi_{m_1}(O) - \phi_{m_2}(O) = p_1(X) - p_2(X) - \Delta m_{12}.$$

This simplification follows because the outcome term $Y$is common to both model-specific signed errors and cancels in their contrast.

Let

$$\phi(O) = \begin{pmatrix} \phi_M(O) \\ \phi_D(O) \\ \phi_L(O) \\ \phi_R(O) \end{pmatrix}.$$

The variance of the asymptotic linear representation may be estimated by the empirical covariance matrix of the estimated influence functions:

$$\hat{\Omega} = \frac{1}{n}\sum_{i=1}^{n} \hat{\phi}(O_i)\hat{\phi}(O_i)^{\top}. \qquad (20)$$

The estimated standard error of an estimator $\theta$is

$$\mathrm{SE}(\hat{\theta}) = \sqrt{\frac{\hat{\Omega}_{\theta\theta}}{n}}. \qquad (21)$$

In first-order regular settings, Wald-type 95% confidence intervals are calculated as $\hat{\theta} \pm 1.96\ \mathrm{SE}(\hat{\theta})$.

Resampling subjects while preserving the pairing of $Y_i$, $p_{1i}$, and $p_{2i}$ may be used to examine sampling variability. However, resampling-based intervals require separate justification in nonregular settings or when finite-sample distributions depart substantially from the relevant asymptotic approximation.

These procedures target the performance contrasts between fixed prediction rules in the specified target population. If the inferential target instead includes the complete model-development algorithm, including model fitting, variable selection, tuning, or prediction generation, the entire development procedure must be repeated within each resampling iteration.

For the mean-bias contrast, ordinary first-order inference becomes degenerate when $\mu_1-\mu Y=0$ and $\mu_2-\mu Y=0$, because the first derivative of the contrast functional vanishes. Wald inference for $\Delta M_{12}$ should therefore be interpreted only in first-order regular settings.

To obtain a confidence interval that remains valid at this singular point, we additionally used the factorization $\Delta M_{12} = ab$, where $a = E\{p_1(X) - p_2(X)\}$ and $b = E\{p_1(X) + p_2(X) - 2Y\}$. A joint Wald confidence ellipse for $(a,b)$ was projected through $g(a,b) = ab$. If the joint ellipse has asymptotic coverage $1 - \alpha$, its image provides a confidence set for $\Delta M_{12}$ with asymptotic coverage at least $1 - \alpha$. This approach retains the paired plug-in point estimate and the exact finite-sample Brier-risk identity, but may be conservative because multiple values of $(a,b)$ correspond to the same product. Details and simulations are provided in Supplementary Methods S1 and Supplementary Table S4. The construction addresses the first-order-degenerate mean-bias contrast and differs from sample-splitting approaches developed for null-degenerate contrasts in nonparametric predictiveness measures.[16]

Normal approximation may also be unreliable for extremely small dispersion or alignment-contribution contrasts whose finite-sample distributions are highly concentrated or skewed. In such settings, first-order Wald intervals should not be interpreted mechanically. The projection procedure described above is specific to the mean-bias contrast and does not address settings in which finite-sample normal approximation is inadequate for dispersion or alignment-contribution contrasts; development of alternative procedures for such settings is beyond the scope of the present study.

### 2.11 Optional proxy-relative model-specific diagnostics

If model-specific diagnostic quantities are nevertheless desired, an additional proxy-risk model may be introduced. Let $\tilde{r}(X)$ denote a flexible estimate or approximation of the conditional event risk, preferably generated using cross-fitting or an independent sample. Proxy-relative components may then be defined as

$$\tilde{M}(p) = \left[E\{p(X) - \tilde{r}(X)\}\right]^2,$$

$$\tilde{D}(p) = \mathrm{Var}\{p(X)\} - \mathrm{Var}\{\tilde{r}(X)\},$$

and

$$\tilde{L}(p) = -2\,\mathrm{Cov}[\tilde{r}(X), p(X) - \tilde{r}(X)]\,.$$

These quantities satisfy

$$E\left[\{p(X) - \tilde{r}(X)\}^2\right] = \tilde{M}(p) + \tilde{D}(p) + \tilde{L}(p).$$

However, this is a decomposition of the squared distance from the proxy risk, $E\left[\{p(X) - \tilde{r}(X)\}^2\right]$, rather than the structural prediction error $E\left[\{p(X) - r(X)\}^2\right]$, unless $\tilde{r}(X) = r(X)$ almost surely.

proxy-relative components depend on the predictor set, functional form, estimation method, tuning procedure, and sample used to construct $\tilde{r}(X)$. They also inherit the proxy model's approximation error and estimation uncertainty. Cross-fitting can reduce overfitting and dependence caused by evaluating the proxy on the data used to fit it, but it does not make the proxy equal to the true conditional risk or eliminate the dependence of these diagnostics on the chosen proxy model.

Proxy-relative components should therefore be interpreted as exploratory diagnostics conditional on the selected proxy-risk model. They should not be presented as identified model-specific decompositions of the population Brier risk.

The primary empirical analysis is consequently based on the exactly identified between-model estimands $\Delta R_{12}$, $\Delta M_{12}$, $\Delta D_{12}$, and $\Delta L_{12} = -2\Delta A_{12}$. Proxy-relative model-specific quantities are optional and should be reported separately from the primary contrast analysis.

## 3. Simulation study

### 3.1 Objectives

Because the Brier-UDAM decomposition is an algebraic identity, the simulation study was not designed merely to confirm that the identity holds. Instead, we conducted two simulation experiments with distinct objectives.

The first, termed the component-interpretation experiment, examined whether the signed decomposition components responded in an interpretable manner to controlled perturbations of a known conditional event risk function. In this experiment, we also evaluated whether the observable contrast estimators were centered on their corresponding replicate-specific risk-based targets and whether their sum reproduced the paired empirical Brier-risk difference exactly within each replicate.

The second, termed the large-sample inference experiment, evaluated the operating characteristics of the influence-function-based variance estimators for population-level contrasts between fixed prediction rules. Specifically, we assessed Monte Carlo bias, empirical sampling variability, estimated standard errors, and coverage of nominal 95% Wald confidence intervals.

The two experiments used the same linear and nonlinear data-generating mechanisms but differed in how the prediction rules and estimands were defined. The component-interpretation experiment is summarized in Table 1; the fixed-rule inference experiment is described separately in Section 3.6, with its operating characteristics reported in Supplementary Tables S1 and S2.

### 3.2 Data-generating mechanisms

For each replicate, seven mutually independent standard normal covariates $X_1$,...,$X_7$ were independently generated from $N(0,1)$. The sixth covariate, $X_6$, and the seventh covariate, $X_7$, were excluded from the outcome model. $X_6$ was used only to construct the rank-misalignment rule, whereas $X_7$ was used only to construct the noise rule.

In the linear setting, the true linear predictor was

$$\eta_{\text{lin}} = -1.0 + 0.8X_1 - 0.6X_2 + 0.5X_3 + 0.4X_4 - 0.3X_5.$$

In the nonlinear setting, the true linear predictor was

$$\eta_{\text{nonlin}} = \eta_{\text{lin}} + 0.6\left(X_1^2 - 1\right) + 0.5X_2X_3.$$

The true conditional event risk was

$$r(X) = \text{expit}(\eta),$$

where $\eta$was the corresponding linear or nonlinear predictor. The binary outcome was generated according to

$$Y|X \sim \text{Bernoulli}\{r(X)\}.$$

For the component-interpretation experiment, we generated $n = 1{,}500$ subjects in each of $1{,}000$ Monte Carlo replicates for each data-generating mechanism. Random seed 20260710 was used. All candidate prediction rules were constructed directly from the known conditional risk, so no model fitting or proxy-risk estimation was required. Within each replicate, all candidate rules were evaluated on the same subjects.

### 3.3 Prediction rules for the component-interpretation experiment

Let $r_i$ denote the true conditional risk for subject $i$, and define the replicate-specific empirical mean conditional risk as

$$\bar{p} = \frac{1}{n}\sum_{i=1}^{n} r_i .$$

Seven candidate prediction rules were considered.

The oracle rule was

$$p_i^{\text{oracle}} = r_i.$$

The mean-shift rule was

$$p_i^{\text{shift}} = \text{clip}(r_i + 0.10) .$$

The compression rule was

$$p_i^{\text{compression}} = \bar{r} + 0.60\left(r_i - \bar{r}\right),$$

which reduced the empirical dispersion of the predicted risks around their mean.

The expansion rule was

$$p_i^{\text{expansion}} = \text{clip}\{\bar{r} + 1.40(r_i - \bar{r})\}$$

which increased the empirical dispersion of the predictions.

The noise rule was

$$p_i^{\text{noise}} = \text{clip}(r_i + 0.05\text{X}_{7i}) .$$

Because $X_7$ was independent of the covariates determining the conditional event risk, this rule introduced subject-level prediction noise without changing the outcome-generating mechanism.

For the rank-misalignment rule, the empirical values of $r_i$were reassigned to subjects according to the ranks of $X_{6i}$. Let $p_i^{\text{mis}}$ denote the reassigned value. This construction preserved the empirical distribution, and therefore the empirical mean and variance, of the conditional event risks exactly, while disrupting the subject-level correspondence between $p_i^{\text{mis}}$ and $r_i$. The resulting mean-bias and dispersion components were therefore zero up to numerical precision, and the perturbation was expressed almost entirely through the risk-alignment contribution.

The combined rule was

$$p_i^{\text{combined}} = \text{clip}\left\{\bar{r} + 1.25(r_i - \bar{r}) + 0.06 + 0.35(p_i^{\text{mis}} - \bar{r})\right\}.$$

The clipping operator was defined as

$$\text{clip}(z) = \min\{1 - 10^{-6}, \max(10^{-6}, z)\}.$$

The candidate rules were selected to generate interpretable perturbation patterns. They were not intended to imply that mean bias, dispersion mismatch, and risk alignment can generally be varied independently. In particular, clipping may simultaneously affect more than one component.

Because several rules in this experiment depended on the replicate-specific quantity $r$, they were not fixed subject-level functions in the population. This experiment targeted replicate-specific finite-sample quantities and focused on component interpretation and observable recovery. The influence-function theory for fixed prediction rules was evaluated in the separate large-sample experiment.

### 3.4 Replicate-specific simulation targets

For each non-oracle candidate rule $p$, we calculated the following replicate-specific quantities defined relative to the true conditional event risks:

$$M_n(p) = \left[\frac{1}{n}\sum_{i=1}^{n} (p_i - r_i)\right]^2,$$

$$D_n(p) = \mathrm{Var}_n(p) - \mathrm{Var}_n(r),$$

and

$$L_n(p) = -2\mathrm{Cov}_n(r, p - r),$$

where empirical variances and covariances were calculated using denominator $n$. Their sum was

$$T_n(p) = M_n(p) + D_n(p) + L_n(p) = \frac{1}{n}\sum_{i=1}^{n} (p_i - r_i)^2.$$

The oracle predictor satisfied

$$M_n(p^{\text{oracle}}) = D_n(p^{\text{oracle}}) = L_n(p^{\text{oracle}}) = T_n(p^{\text{oracle}}) = 0.$$

Consequently, these model-specific quantities were numerically identical to the corresponding replicate-specific contrasts between the candidate rule and the oracle predictor. We therefore defined

$$\Delta M_n^{\text{oracle}}(p) = M_n(p),$$

$$\Delta D_n^{\text{oracle}}(p) = D_n(p),$$

$$\Delta L_n^{\text{oracle}}(p) = L_n(p),$$

and

$$\Delta R_n^{\text{oracle}}(p) = T_n(p).$$

These quantities served as the replicate-specific simulation targets against which the observable contrast estimators were evaluated.

### 3.5 Observable estimators and Monte Carlo performance measures

For each candidate rule $p$, the observable component contrasts relative to the oracle predictor were calculated from the realized outcomes and paired predictions as

$$\widehat{\Delta M}_n(p) = \left(\bar{p} - \bar{Y}\right)^2 - \left(\bar{r} - \bar{Y}\right)^2,$$

$$\widehat{\Delta D}_n(p) = \text{Var}_n(p) - \text{Var}_n(r),$$

and

$$\widehat{\Delta L}_n(p) = -2\text{Cov}_n(Y,p - r).$$

The empirical Brier-risk contrast was

$$\widehat{\Delta R}_n(p) = \frac{1}{n}\sum_{i=1}^{n}\left[(Y_i - p_i)^2 - (Y_i - r_i)^2\right].$$

For each replicate, the exact finite-sample identity

$$\widehat{\Delta R}_n(p) = \widehat{\Delta M}_n(p) + \widehat{\Delta D}_n(p) + \widehat{\Delta L}_n(p)$$

was evaluated numerically.

For component $C \in \{M,D,L,R\}$, the estimation error in replicate $m$was defined as

$$E_{C,m} = \widehat{\Delta C}_{n,m} - \Delta C_{n,m}^{\text{oracle}}.$$

Monte Carlo performance was summarized by the mean estimation error,

$$\text{MCBias}_C = \frac{1}{m}\sum_{m=1}^{m} E_{C,m},$$

and by the 2.5th and 97.5th percentiles of the replicate-specific estimation errors.

The dispersion estimator equaled its replicate-specific risk-based target exactly within every replicate because both were calculated as

$$\text{Var}_n(p) - \text{Var}_n(r).$$

In contrast, the mean-bias and alignment estimators depended on the realized binary outcomes and therefore retained finite-sample variability around their risk-based targets.

**3.6 Large-sample inference experiment**

A separate simulation experiment was conducted to evaluate inference for population-level contrasts between fixed prediction rules. In this experiment, each prediction rule was defined as a fixed function of the corresponding subject's augmented covariate vector and conditional event risk. The resulting $(Y_i,P_{1i},P_{2i})$, $i = 1,\ldots,n$, were independent and identically distributed, as assumed in the influence-function derivation in Section 2.10.

Sample sizes of $n = 250, 500, and\ 1{,}500$ were considered under both the linear and nonlinear data-generating mechanisms. For each combination of setting, sample size, and prediction rule, 1,000 Monte Carlo replicates were generated.

Let $\mu_r = E\{r(X)\}$ denote the population mean conditional event risk. In this experiment, the compression, expansion, and combined prediction rules were defined using $\mu_r$, rather than the replicate-specific empirical mean $r$. This ensured that each prediction was a fixed subject-level function and did not depend on other observations in the evaluation sample.

The compression rule was

$$p^{\text{compression}}(X) = \mu_r + 0.60\{r(X) - \mu_r\},$$

and the expansion rule was

$$p^{\text{expansion}}(X) = \text{clip}\left[\mu_r + 1.40\{r(X) - \mu_r\}\right].$$

The combined rule was defined analogously as

$$p^{\text{combined}}(X) = \text{clip}\left[\mu_r + 1.25\{r(X) - \mu_r\} + 0.06 + 0.35\{p^{\text{mis}}(X) - \mu_r\}\right].$$

For the population-level rank-misalignment rule, the independent covariate $X_6$was transformed as

$$p^{\text{mis}}(X) = F_r^{-1}\{\Phi(X_6)\},$$

where $F_r$denotes the marginal cumulative distribution function of $r(X)$, $F_r^{-1}$ its generalized inverse, and $\Phi$the standard normal cumulative distribution function. Because $X_6$ was independent of the covariates determining $r(X)$,

$$p^{\text{mis}}(X) \sim r(X),$$

where $\sim$ denotes equality in distribution, while $p^{\text{mis}}(X)$ was independent of $r(X)$. The rule preserved the population mean and variance of the conditional-risk distribution while disrupting subject-level risk alignment.

Population contrast targets were approximated using an independent reference sample of size $N_{\text{ref}} = 1{,}000{,}000$.

For each component and for the total Brier-risk contrast, the estimators in Section 2.9 and the influence-function-based standard errors in Section 2.10 were calculated.

For a generic target $\theta$, Monte Carlo bias was defined as

$$\text{Bias} = \frac{1}{m}\sum_{m=1}^{m}(\hat{\theta}_m - \theta_{\text{ref}}),$$

and the empirical Monte Carlo standard deviation was

$$\mathrm{SD}_{\mathrm{MC}} = \left[ \frac{1}{m-1} \sum_{m=1}^{m} (\hat{\theta}_m - \bar{\hat{\theta}})^2 \right]^{1/2}.$$

The mean estimated standard error was

$$\mathrm{SE} = \frac{1}{m} \sum_{m=1}^{m} \mathrm{SE}_m .$$

The standard-error ratio was defined as

$$SE\ ratio = \frac{\mathrm{SE}}{\mathrm{SD}_{\mathrm{MC}}}.$$

Values below 1 indicated underestimation of sampling variability, whereas values above 1 indicated overestimation.

Coverage was defined as the proportion of nominal 95% Wald confidence intervals $\hat{\theta}_m \pm 1.96\mathrm{SE}_m$ that contained $\theta_{\mathrm{ref}}$.

The mean-bias contrast required special consideration. Writing $\Delta M = (\mu_1 - \mu_Y)^2 - (\mu_2 - \mu_Y)^2$, where $\mu_j = E\{p_j(X)\}$, $\mu_Y = E(Y)$, its gradient with respect to $(\mu_1, \mu_2, \mu_Y)$ is

$$\nabla \Delta M = \begin{pmatrix} 2(\mu_1 - \mu_Y) \\ -2(\mu_2 - \mu_Y) \\ -2(\mu_1 - \mu_2) \end{pmatrix}.$$

The ordinary first-order delta method is degenerate when $\mu_1 = \mu_2 = \mu_Y$, because the gradient then vanishes. Mean-bias scenarios were therefore classified as first-order regular or first-order degenerate. Wald coverage in degenerate settings was reported descriptively but was not interpreted as evidence supporting the validity of ordinary first-order inference.

For dispersion and alignment-contribution contrasts generated only by rare probability clipping, the first-order influence functions were not algebraically degenerate. Nevertheless, the corresponding finite-sample distributions could be highly concentrated near zero or markedly skewed, resulting in slow convergence to the normal approximation and underestimation of sampling variability.

## 3.7 Results

### 3.7.1 Component patterns and recovery of replicate-specific targets

The mean event proportion was approximately 0.314 under both data-generating mechanisms. The risk-based component patterns were consistent with the intended perturbations (Table 2 and Figure 2).

The mean-shift rule was dominated by the mean-bias component. Small dispersion and alignment contributions remained because clipping altered some of the shifted probabilities, particularly in the nonlinear setting (Table 2 and Figure 2).

The compression rule produced a negative dispersion contribution and a larger positive alignment contribution (Table 2 and Figure 2). Conversely, the expansion rule produced a positive dispersion contribution and a negative alignment contribution. These opposing contributions were expected because altering the spread of the predictions changes both $\mathrm{Var}\{p(X)\}$ and $\mathrm{Cov}\left[r(X), p(X) - r(X)\right]$. These opposing movements reflect the structural coupling between dispersion and alignment.

The noise rule primarily increased the dispersion component, with a smaller alignment contribution arising partly from probability clipping. The empirical rank-misalignment rule provided the clearest finite-sample separation of alignment from mean bias and dispersion. Because it preserved the empirical distribution of the conditional event risks exactly, its mean-bias and dispersion components were zero up to numerical precision. Its alignment contribution and total structural error were approximately 0.092 in the linear setting and 0.119 in the nonlinear setting. The combined rule produced simultaneous mean-bias, dispersion, and alignment contributions (Table 2 and Figure 2).

The observable component contrasts were centered near their replicate-specific risk-based targets (Figure 3). The dispersion contrast coincided exactly with its target within every replicate because the variance of the conditional event risk cancelled algebraically. The mean-bias and alignment-contribution contrasts retained finite-sample variability because their observable representations replaced the conditional event risk with the realized binary outcome (Figure 3). Across the component-interpretation experiment, the largest absolute Monte Carlo mean estimation error was 0.000121; the corresponding Monte Carlo standard error was 0.000077, giving an absolute standardized Monte Carlo error of 1.56. The observed deviation was small relative to Monte Carlo uncertainty.

The identity $\widehat{\Delta R}_n = \widehat{\Delta M}_n + \widehat{\Delta D}_n + \widehat{\Delta L}_n$ held to machine precision in every replicate. The largest absolute identity residual was approximately $4.16 \times 10^{-17}$.

### 3.7.2 Operating characteristics of large-sample inference

Coverage and standard-error calibration are summarized in Supplementary Table S1, complete numerical results are provided in Supplementary Table S2, and coverage patterns across sample sizes are displayed in Supplementary Figure S1. In the separate large-sample inference experiment, the total Brier-risk contrast generally showed satisfactory first-order behavior. Across all examined settings, coverage of the nominal 95% Wald confidence interval for the total contrast ranged from 0.930 to 0.960. The mean influence-function-based standard error was also close to the empirical Monte Carlo standard deviation.

For dispersion and alignment-contribution contrasts of practically non-negligible magnitude, Wald coverage was generally close to the nominal level. Across these regular settings, median coverage was 0.947 for the dispersion contrast and 0.946 for the alignment-contribution contrast (Supplementary Tables S1 and S2; Supplementary Figure S1).

An important exception occurred for the very small dispersion and alignment-contribution contrasts generated by rare clipping under the linear mean-shift rule. For the dispersion contrast, coverage was 0.572, 0.749, and 0.869 at sample sizes of 250, 500, and 1,500, respectively. For the alignment-contribution contrast, the corresponding coverage probabilities were 0.556, 0.744, and 0.868 (Supplementary Tables S1 and S2; Supplementary Figure S1). In these settings, the mean estimated standard error was smaller than the empirical Monte Carlo standard deviation, indicating slow and non-Gaussian finite-sample behavior despite the absence of algebraic first-order degeneracy.

For the mean-bias contrast, algebraic first-order regularity required a nonzero gradient of the contrast functional. When both prediction rules had zero signed mean error, the first-order derivative vanished and Wald coverage was essentially 1.000, showing severe conservatism; this did not validate ordinary nominal first-order inference. Some formally regular settings with extremely small mean-bias contrasts also showed substantial overcoverage, indicating that a nonzero gradient alone does not guarantee an accurate finite-sample normal approximation (Supplementary Tables S1 and S2; Supplementary Figure S1).

In the additional projection-inference experiment, the projected intervals had empirical coverage from 0.981 to 1.000 across exactly degenerate, local-to-zero, one-factor-null, and regular settings at sample sizes of 250, 500, and 1,500 (Supplementary Table S4). The projection intervals were conservative, particularly near the singular intersection $a = b = 0$, where either factor being zero implies $\Delta M_{12} = 0$. In the exactly degenerate setting, the median projection-interval width decreased from 0.001329 at n=250 to 0.000219 at n=1,500, consistent with the second-order $n^{-1}$ scale of the product near the origin.

Overall, these findings support the influence-function variance estimator for regular component contrasts of non-negligible magnitude and for the total Brier-risk contrast. For the mean-bias contrast, projection of the joint confidence ellipse provides an asymptotically valid, though potentially conservative, alternative when the first-order gradient vanishes or is small. These findings do not support routine Wald confidence-interval inference for the clipping-induced near-zero dispersion and alignment-contribution contrasts that exhibited poor finite-sample normal approximation in the examined settings. These contrasts were not algebraically first-order degenerate, and the results should not be interpreted as implying that near-zero dispersion or alignment contrasts are generally nonregular. A general procedure for settings with inadequate finite-sample normal approximation is not developed here.

## 4. Illustrative empirical analysis

### 4.1 Aim

We applied the observable paired-contrast decomposition to a large public clinical dataset to illustrate how Brier-UDAM can be used when several fitted prediction rules have similar overall Brier risks but differ in their underlying prediction-error structure. The empirical analysis was designed as a methodological illustration rather than as the development or endorsement of a clinical readmission model.

Three prediction algorithms were compared: standard logistic regression (GLM), LASSO-penalized logistic regression, and random forest. All algorithms were developed using the same prespecified predictor information and were evaluated on the same held-out subjects. Conventional measures of predictive performance were examined first, followed by the paired Brier-UDAM contrasts. The purpose was to determine whether small differences in aggregate Brier risk concealed larger and opposing mean-bias, dispersion, or alignment contributions.

Unlike the previous simulation experiments, the conditional event risk was unknown in this application. We therefore did not estimate model-specific latent-risk UDAM components or introduce a proxy-risk model. The empirical analysis used only the exactly observable paired contrasts $\Delta M$, $\Delta D$, $\Delta L = -2\Delta A$, and $\Delta R$.

### 4.2 Data source, cohort, and outcome

The empirical illustration used the current public release of the UCI Diabetes 130-US Hospitals for Years 1999–2008 dataset[17,18]. The database contains hospital encounters for patients with diabetes treated at 130 US hospitals and integrated delivery networks during 1999–2008.

Because some patients had multiple recorded hospital encounters, we retained one encounter per patient to obtain independent observational units. For each patient, we retained the encounter with the smallest encounter identifier as a deterministic operational representation of the first recorded encounter. Patients whose retained encounter ended in death or discharge to hospice were excluded. The binary outcome was inpatient readmission within 30 days after discharge, defined by the database category readmitted = "<30".

The current public dataset contained 101,766 encounters among 71,518 unique patients. After retaining the first encounter for each patient and excluding death or hospice discharges, the operational analysis cohort contained 69,973 patients, of whom 6,277 (9.0%) experienced readmission within 30 days. Application of these prespecified criteria to the current public release did not exactly reproduce the historical analytic counts reported for the original dataset; we therefore used the transparently defined cohort obtained from the current public file rather than imposing additional undocumented exclusions. The operational cohort passed prespecified checks confirming one retained encounter per patient, absence of retained death or hospice discharges, and complete binary outcome ascertainment.

The prediction time was discharge. Candidate predictors were restricted to information available by that time. The prespecified predictor set comprised demographic characteristics (race, sex, and age), admission characteristics (admission type and source), discharge disposition, medical specialty, primary diagnosis category, length of stay, numbers of laboratory procedures, procedures and medications, prior outpatient, emergency, and inpatient encounters, number of diagnoses, maximum serum glucose category, HbA1c result category, insulin treatment, medication-change status, and use of diabetes medication. Variables with substantial missingness, including weight and payer code, were not used. Missing or unknown values for retained categorical predictors were represented by prespecified explicit categories rather than being excluded.

### 4.3 Model development and held-out validation

Using random seed 20260822, the operational cohort was divided by simple random sampling, independently of outcome status, into a development sample containing approximately two thirds of the patients and a common held-out validation sample containing the remaining one third. The development sample contained 46,648 patients and the validation sample contained 23,325 patients. All model fitting and hyperparameter selection were restricted to the development sample, and the validation outcomes were not used for model development or tuning. All preprocessing rules and predictor definitions were fixed before model fitting. The same predictor set was supplied to all three algorithms.

The GLM was a logistic regression model containing the prespecified predictors as main effects. The LASSO model was fitted using logistic regression with an $L_1$ penalty. The penalty parameter was selected by five-fold internal cross-validation within the development sample using binomial deviance, with the value minimizing cross-validated deviance selected for the final model.

The random forest was tuned by five-fold internal cross-validation within the development sample. Candidate values of the number of variables considered at each split were based on approximately the square root, one third, and one half of the number of prespecified predictors, and candidate minimum terminal-node sizes were 5, 20, and 50. Hyperparameters were selected by mean validation log loss. Three hundred trees were used for the cross-validation fits and 1,000 trees for the final fitted random forest. No class weighting, outcome over-sampling, under-sampling, or synthetic minority over-sampling was used, because the analysis targeted probability prediction rather than classification.

After all model-development decisions had been completed, each fitted model generated a predicted 30-day readmission probability for every subject in the common held-out validation sample. This produced subject-level paired comparisons between algorithms. Numerical probability bounding to $[10^{-6}, 1 - 10^{-6}]$ was used where required for stable probability-scale or logit calculations.

**4.4 Statistical analysis**

Conventional predictive performance was evaluated before examining the Brier-UDAM decomposition. For each fitted rule, we calculated the Brier score, area under the receiver operating characteristic curve (AUC), calibration intercept, and calibration slope. Flexible calibration curves were additionally estimated using a spline model on the logit prediction scale, and the distributions of predicted risks were examined graphically (Supplementary Figure S2). The Brier-score confidence interval was based on the empirical variability of the individual squared prediction errors, the AUC confidence interval used the DeLong method[19], and calibration-intercept and calibration-slope intervals used the corresponding logistic-regression large-sample approximations.

The primary Brier-UDAM analysis compared the three fitted prediction rules pairwise: GLM minus LASSO, GLM minus random forest, and LASSO minus random forest. Contrasts were defined throughout as the first prediction rule minus the second. A positive ΔR indicated a larger Brier risk, and therefore poorer overall probabilistic prediction performance, for the first rule.

For each comparison, the observable estimators described in Section 2.9 were used:

$$\widehat{\Delta M} = \left(\bar{p}_1 - \bar{Y}\right)^2 - \left(\bar{p}_2 - \bar{Y}\right)^2,$$

$$\widehat{\Delta D} = Var_n(p_1) - Var_n(p_2),$$

$$\widehat{\Delta L} = -2Cov_n\{Y, p_1 - p_2\},$$

and

$$\widehat{\Delta R} = \frac{1}{n}\sum_{i=1}^{n}\left[(Y_i - p_{1i})^2 - (Y_i - p_{2i})^2\right].$$

All empirical variances and covariances used the denominator-n convention. The analysis program required

$$\widehat{\Delta R} = \widehat{\Delta M} + \widehat{\Delta D} + \widehat{\Delta L}$$

to hold to within $10^{-12}$ for every comparison.

Because model development was performed in a separate development sample, inference in the held-out sample treated the three fitted prediction functions as fixed rules and quantified sampling uncertainty in their validation-population performance. Influence-function-based standard errors and nominal 95% Wald confidence intervals were calculated as described in Section 2.10.

For the mean-bias contrast, we additionally calculated the projection confidence interval described in Supplementary Methods S1. This analysis was included as a sensitivity analysis because $\Delta M$ can approach the first-order-degenerate point when both fitted rules have negligible signed mean error. The estimated norm of the first-order gradient was reported together with the ordinary Wald and projection intervals.

The empirical analysis was intended to illustrate the interpretation of the decomposition. Statistical significance of individual components was not used to rank or select the algorithms. Dispersion and alignment contributions were considered together with the total Brier-risk contrast.

### 4.5 Empirical results

#### 4.5.1 Conventional predictive performance

Predictive performance in the held-out validation sample was similar across the three algorithms in terms of overall Brier risk and discrimination (Table 3). The Brier score was 0.07991 (95% CI, 0.07700–0.08282) for GLM, 0.07984 (0.07692–0.08275) for LASSO, and 0.08009 (0.07722–0.08295) for random forest. The corresponding AUCs were 0.620 (0.607–0.632), 0.621 (0.608–0.633), and 0.618 (0.606–0.630), respectively.

Calibration differed more clearly between the algorithms. Calibration intercepts were −0.022 (−0.068 to 0.024) for GLM and −0.022 (−0.067 to 0.024) for LASSO, whereas random forest had a calibration intercept of −0.066 (−0.111 to −0.020). Calibration slopes were 0.796 (0.710–0.883) for GLM, 0.883 (0.792–0.975) for LASSO, and 0.713 (0.636–0.791) for random

forest. The flexible calibration curves and predicted-risk distributions likewise showed differences in the probability distributions produced by the three fitted rules despite their similar aggregate Brier scores (Supplementary Figure S2).

### 4.5.2 Brier-UDAM paired contrasts

The Brier-UDAM decomposition showed that the small differences in overall Brier risk concealed substantially larger and opposing dispersion and alignment contributions (Table 4 and Figure 4).

For GLM versus LASSO, the total Brier-risk contrast was $\widehat{\Delta R}$=0.0000714 (95% CI, 0.0000062 to 0.0001366). The mean-bias contrast was essentially zero, $\widehat{\Delta M}$=0.0000001 (−0.0000003 to 0.0000004). GLM had greater prediction dispersion than LASSO, $\widehat{\Delta D}$ =0.0002373 (0.0002209 to 0.0002538), but this was substantially offset by a negative alignment contribution, $\widehat{\Delta L}$ =−0.0001660 (−0.0002316 to −0.0001005). The opposing alignment contribution made the total Brier-risk difference substantially smaller than the dispersion contrast.

For GLM versus random forest, the total Brier-risk contrast was $\widehat{\Delta R}$= −0.0001783 (−0.0004556 to 0.0000991). The mean-bias contrast was small, $\widehat{\Delta M}$= −0.0000249 (−0.0000507 to 0.0000009). GLM had lower prediction dispersion than random forest, $\widehat{\Delta D}$ = −0.0004928 (−0.0005960 to −0.0003895), whereas the alignment contribution was positive, $\widehat{\Delta L}$ = 0.0003394 (0.0000487 to 0.0006302). The opposing dispersion and alignment contributions therefore yielded a considerably smaller aggregate Brier-risk contrast.

The strongest compensation was observed for LASSO versus random forest. The total Brier-risk contrast was $\widehat{\Delta R}$=−0.0002497 (−0.0005121 to 0.0000128). The mean-bias contrast was again small, $\widehat{\Delta M}$=−0.0000250 (−0.0000509 to 0.0000010). LASSO had substantially lower prediction dispersion than random forest, $\widehat{\Delta D} =$ −0.0007301 (−0.0008276 to −0.0006327), whereas its alignment contribution was positive, $\widehat{\Delta L}$ =0.0005055 (0.0002303 to 0.0007806). Thus, the two principal signed contributions acted strongly in opposite directions, leaving an aggregate Brier-risk difference much smaller than either component.

Across the three comparisons, the mean-bias contributions were negligible relative to the dispersion and alignment contributions. The empirical results therefore illustrate that prediction rules with similar aggregate Brier risks can nevertheless differ substantially in prediction spread and risk-error alignment, with the aggregate Brier-risk contrast reflecting partial cancellation between these signed components.

### 4.5.3 Mean-bias projection sensitivity analysis

The GLM–LASSO comparison was closest to the first-order-degenerate region (gradient norm, 0.00495), for which the projection interval was wider than the ordinary Wald interval. The other two comparisons were farther from the singular point and showed more similar Wald and projection intervals (Supplementary Table S3). These results illustrate the potential relevance of projection inference when the mean-bias contrast approaches its first-order-degenerate region.

## 5. Discussion

### 5.1 Principal findings

Brier-UDAM decomposes a paired Brier-risk contrast into mean-bias, dispersion, and alignment contributions while retaining a latent-risk interpretation. Model-specific dispersion and alignment depend on the conditional event risk, but their between-model contrasts admit exact observed-data representations. The complete paired decomposition can be evaluated from observed binary outcomes and paired predictions without estimating the conditional-risk function or introducing a proxy model.

The simulation studies supported both the interpretation and estimation of the paired Brier-UDAM contrasts. The observable estimators recovered their corresponding risk-based targets, and their contributions reproduced the paired empirical Brier-risk difference exactly. Influence-function-based inference performed satisfactorily for the total contrast and for regular component contrasts with adequate finite-sample behavior. For the mean-bias contrast, the projection procedure provided an asymptotically valid alternative when ordinary first-order inference degenerates, although it may be conservative near the singular point. These results distinguish the exact algebraic decomposition from the separate question of sampling inference.

The public-data illustration showed that similar held-out Brier risks can coexist with much larger dispersion and alignment differences acting in opposite directions. The aggregate Brier-risk contrast was small because these contributions partially cancelled. This diagnostic separation is the main practical role of Brier-UDAM.

### 5.2 Interpretation of the signed components

The dispersion and alignment contributions are signed accounting terms that describe related aspects of the variance of prediction error. Changes in the spread of predicted probabilities can also change how subject-level prediction errors vary across the underlying risk distribution, producing opposing component movements and partial cancellation. For example, compression of predictions toward their mean produces a negative dispersion component because the predictions become less variable than the conditional risks, while higher-risk individuals tend to be underpredicted and lower-risk individuals tend to be overpredicted, producing a positive alignment contribution. Expansion produces the opposite pattern.

The distinction remains informative despite this partial offsetting. The dispersion component characterizes the marginal spread of the predicted probabilities, whereas the alignment contribution characterizes the subject-level correspondence between prediction errors and underlying risks. The rank-misalignment perturbation illustrates the difference: reassignment of the same risk values across individuals preserved their empirical mean and variance, leaving the mean-bias and dispersion components approximately zero, while the loss of subject-level correspondence was expressed almost entirely through the alignment contribution. Separating dispersion and alignment exposes distinct error structures that can cancel within the total Brier-risk difference.

Their structural coupling means that the sign of either component alone is insufficient to establish better or worse predictive performance. A negative alignment contribution, for example, may partly offset excessive prediction dispersion. Interpretation should account for the pair of contributions, their sum, $D(p) + L(p) = \mathrm{Var}\{p(X) - r(X)\}$, and the total Brier-risk contrast.

The alignment-contribution contrast also does not localize prediction error within the risk distribution. It summarizes a second-order association between outcomes and subject-level prediction differences but does not identify whether discrepancies occur primarily among low-risk or high-risk individuals. Calibration curves, residual analyses, and subgroup-specific assessments therefore remain necessary for more detailed diagnostic evaluation. Similarly, a small mean-bias contrast does not imply that either model has little mean bias, because the two models may have similar absolute deviations from the event prevalence.

The nested-information result provides a benchmark for interpreting these opposing component movements. If each successive prediction equals the conditional event risk given an expanded information set, mean bias remains zero, prediction dispersion increases toward the dispersion of the true risks, and the Brier-scale alignment contribution decreases by twice the increase in dispersion. The resulting net reduction in Brier risk equals the increase in prediction dispersion. In this ideal setting, greater dispersion represents recovery of genuine risk heterogeneity rather than deterioration.

Empirical model sequences may depart from this benchmark because of model misspecification, finite-sample estimation, regularization, or overfitting. An increase in dispersion without a sufficiently large improvement in alignment indicates that predictions have become more spread out without recovering a corresponding amount of true risk structure. Changes in mean bias additionally indicate whether the average prediction level has corrected or introduced systematic overprediction or underprediction.

The linear prediction-path result adds a local interpretation. Along a path connecting two arbitrary prediction functions, alignment changes linearly and dispersion changes quadratically. When the mean prediction is preserved, an interior Brier-optimal point occurs where the marginal alignment gain is balanced by the marginal dispersion change. This clarifies how models with similar overall Brier risks can occupy different positions in the mean-bias, dispersion, and alignment structure.

The mean-bias contrast does not itself reveal whether either model overpredicts or underpredicts on average, because it is a difference between squared signed mean errors. Model-specific signed mean errors should therefore be reported when directional interpretation is relevant.

### 5.3 Relation to existing performance measures

Brier-UDAM and Murphy-type prediction-conditioned decompositions address different questions.[7–12] Murphy-type decompositions evaluate reliability and resolution conditional on issued predictions, whereas Brier-UDAM explains a paired Brier-risk difference through mean bias, prediction dispersion, and alignment. The algebraic basis of the observable Brier-UDAM contrast is shared with established Yates-type moment decompositions.[13,15] This shared algebraic basis is used in Brier-UDAM to provide latent-risk interpretation and inference for paired contrasts. Calibration and sharpness frameworks address additional features of probabilistic forecast performance.[14] Calibration intercepts and slopes assess specified recalibration relationships, commonly on the logit scale, while UDAM contributions are moment-based quantities on the probability scale. The AUC assesses ranking rather than probabilistic accuracy.[20]

No single measure substitutes for the others. In particular, Brier-UDAM does not establish conditional calibration, localize error over the prediction range, or quantify information loss from compressing a richer predictor set into an issued probability. Calibration curves, calibration intercepts and slopes, discrimination measures, information-set analyses, and the overall Brier risk remain necessary for their respective questions.

### 5.4 Implications for model refinement

Brier-UDAM is a diagnostic framework. When an existing prediction rule is compared with a modified rule using paired out-of-sample predictions, the decomposition can show whether the resulting change in Brier risk is associated primarily with the average prediction level, prediction dispersion, or alignment. Such information may help characterize the effect of recalibration, regularization, or other model modifications. The components are not optimization targets; candidate revisions should ultimately be judged by their overall predictive performance in appropriate out-of-sample data.

### 5.5 Limitations

Several limitations concern the estimands and the information summarized by the decomposition. Model-specific D and A depend on the conditional event-risk function r(X) and cannot be identified from the reduced prediction-outcome distribution alone; estimation from the full observed-data law requires estimation of r(X). Exact reduced-data identification applies to paired between-model contrasts in a common target population. Brier-UDAM also uses only first and second moments and does not characterize the full joint distribution of risks and predictions. Its estimands are population specific and may vary with prevalence, predictor distribution, and case mix.

Inference requires separate caution. The proposed confidence intervals use asymptotic approximations. Ordinary Wald inference may be inadequate when the mean-bias contrast is first-order degenerate or when finite-sample normal approximation is poor. The projection procedure addresses the former case but is specific to the mean-bias contrast and may be conservative. The exact decomposition identity and validity of the point estimators should be distinguished from the finite-sample performance of a particular inferential procedure. Assessment of generalization performance also requires out-of-sample prediction. Reuse of the same data for model fitting and evaluation may produce optimistic estimates, and evaluation of the complete model-development algorithm requires fitting, tuning, selection, and prediction generation to be repeated within the validation or resampling procedure.[21,22]

The empirical illustration adds two practical limitations. It used a single random development-validation split, so the reported confidence intervals quantify sampling uncertainty for the fitted prediction rules in the held-out validation population conditional on the development-stage fits; they do not incorporate uncertainty from repeating the complete model-development procedure. The analysis was designed to demonstrate the paired decomposition and was not a clinical development or validation study of 30-day readmission. The prespecified predictor set, algorithms, and tuning procedures should therefore not be read as evidence of clinical superiority for any evaluated algorithm.

## 6. Conclusion

Paired Brier-risk differences can be decomposed into observable mean-bias, dispersion, and alignment contributions while retaining a latent-risk interpretation. Brier-UDAM formalizes this link to established Yates-type moment identities, shows why paired contrasts can be estimated without modeling the conditional-risk function, and preserves exact finite-sample accounting. Influence-function methods provide inference in regular settings, with projection inference available for the first-order-degenerate mean-bias contrast. The framework adds a diagnostic layer to prediction-rule comparison by showing how distinct error structures combine to produce the observed Brier-risk difference.

## Statements and Declarations

### Ethical considerations

According to the applicable institutional policy, ethics committee review was not required because this methodological study used only simulated data and a publicly available, de-identified dataset and involved no participant recruitment, intervention, or access to identifiable private information.

### Consent to participate

Not applicable. This study involved no direct participation of human subjects and used only simulated data and publicly available, de-identified data.

### Consent for publication

Not applicable. No identifiable personal information, images, or individual case details are presented.

### Declaration of conflicting interest

The authors declare that there are no conflicts of interest with respect to the research, authorship, or publication of this article.

### Funding statement

This work was supported by JSPS KAKENHI Grant Number JP25K13392.

### Data Availability Statement

No new participant-level data were collected for this study. The empirical illustration used the publicly available Diabetes 130-US Hospitals for Years 1999–2008 dataset from the UCI Machine Learning Repository (DOI: 10.24432/C5230J). The reproducibility materials include the analysis code used to construct the operational cohort, generate the development-validation split, fit the prediction models, and reproduce all empirical tables and figures. The dataset is publicly available from the UCI Machine Learning Repository under its stated license.

### Use of generative artificial intelligence

OpenAI ChatGPT (GPT-5-series models, including GPT-5.6 Sol; accessed July 9–August 26, 2026) was used during manuscript preparation to assist with English-language editing, organization and drafting of explanatory text, mathematical and statistical exposition, and development and review of R code used for numerical examples and reproducible analyses. AI assistance was used across the Abstract, Introduction, Methods, examples, Discussion, Supporting Information, and computational reporting. The authors independently reviewed and revised all AI-assisted content, verified mathematical derivations and references, executed and tested the R code, checked numerical results, and made all substantive scientific and editorial decisions. No identifiable, confidential, or proprietary human-subject data were entered into the AI system; inputs were limited to manuscript text, equations, R code, simulated data, publicly available information, and non-identifiable aggregate outputs.

**Figure Legends**

**Figure 1. From the latent-risk representation to the observable paired Brier-UDAM decomposition.**

Model-specific dispersion and alignment quantities are defined relative to the conditional event risk $r(X)$. For two prediction rules evaluated in the same target population, subtraction eliminates the common uncertainty and conditional-risk variance terms, while $Cov\{r(X),p_1 - p_2\} = Cov\{Y,p_1 - p_2\}$. The paired Brier-risk contrast $\Delta R = \Delta M + \Delta D + \Delta L$ can be evaluated from the observed outcome and paired predictions without estimating $r(X)$.

**Figure 2. Oracle Brier-UDAM contributions under controlled perturbations.**

Bars show Monte Carlo means of the mean-bias contribution M, dispersion contribution D, and Brier-scale alignment contribution L=-2A, where $A$ denotes the alignment term. The mean-shift perturbation was dominated by M. Compression and expansion generated D and L contributions with opposing signs, whereas rank misalignment preserved the empirical mean and variance of the conditional event risks and therefore affected almost exclusively L. The opposing signs of D and L reflect the structural coupling between dispersion and alignment.

**Figure 3. Finite-sample recovery of Brier-UDAM contribution targets.**

Points show Monte Carlo mean estimation errors, and vertical ranges show the 2.5th and 97.5th percentiles across 1,000 replicates. Estimation error was the observable contrast minus its target based on known conditional event risks. Errors were centered near zero. The dispersion contrast coincided exactly with its target within each replicate, whereas $M$, $L = -2A$, and total contrasts retained finite-sample variation through realized binary outcomes. Here $M$, $D$, and $L = -2A$ denote the mean-bias, dispersion, and Brier-scale alignment contributions, respectively, and $A$ denotes the alignment term.

**Figure 4. Observable Brier-UDAM decomposition of paired held-out Brier-risk contrasts.**

Bars show the sequential contributions of mean bias ($\Delta M$), dispersion ($\Delta D$), and Brier-scale alignment ($\Delta L = -2\Delta A$) to the paired Brier-risk contrast. Each bar begins at the cumulative value after the preceding contribution, so that the waterfall terminates at $\Delta M + \Delta D + \Delta L = \Delta R$. The black point shows the directly estimated total Brier-risk contrast, with its nominal 95% influence-function-based confidence interval. Contrasts are defined as the first model minus the second. The vertical scale is expressed in units of $10^{-4}$.

**Model-specific latent-risk representation**

$R(p_j) = U + M_j + D_j + L_j$

$D_j$ and $L_j$ are defined relative to $r(X)$

Structural interpretation

→

**Paired subtraction**

$U$ and $Var(r(X))$ cancel

$Cov(r(X), p_1 - p_2) = Cov(Y, p_1 - p_2)$

No estimation of $r(X)$ required

→

**Observable paired decomposition**

$\Delta R = \Delta M + \Delta D + \Delta L$

$M$: mean-bias contribution
$D$: prediction-spread difference
$L$: Brier-scale alignment contribution

Uses paired $Y$, $p_1$, and $p_2$ only

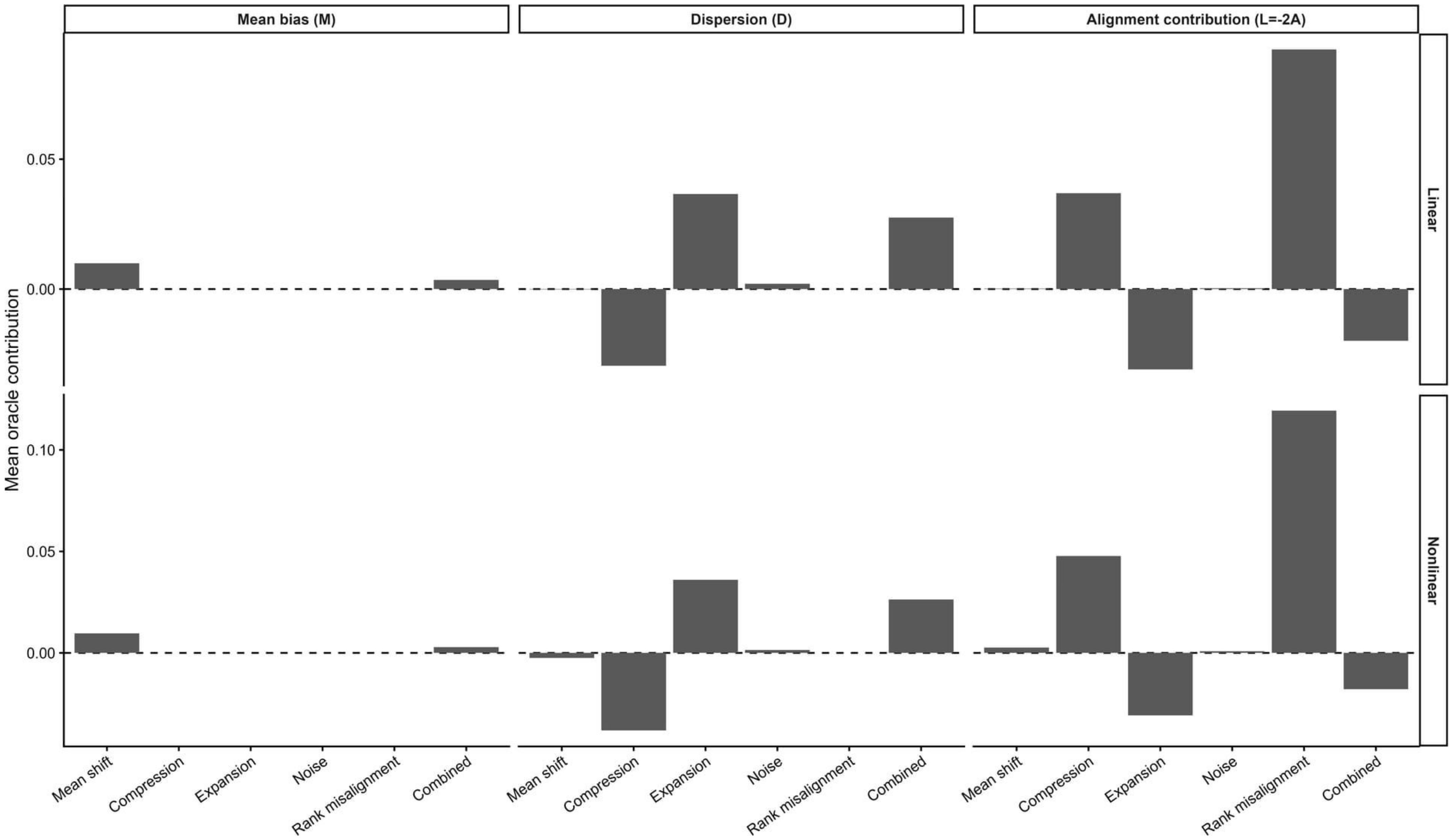

Mean bias (M)
Dispersion (D)
Alignment contribution (L=-2A)
Linear
Nonlinear
Mean oracle contribution
0.05
0.00
0.10
0.05
0.00
Mean shift
Compression
Expansion
Noise
Rank misalignment
Combined

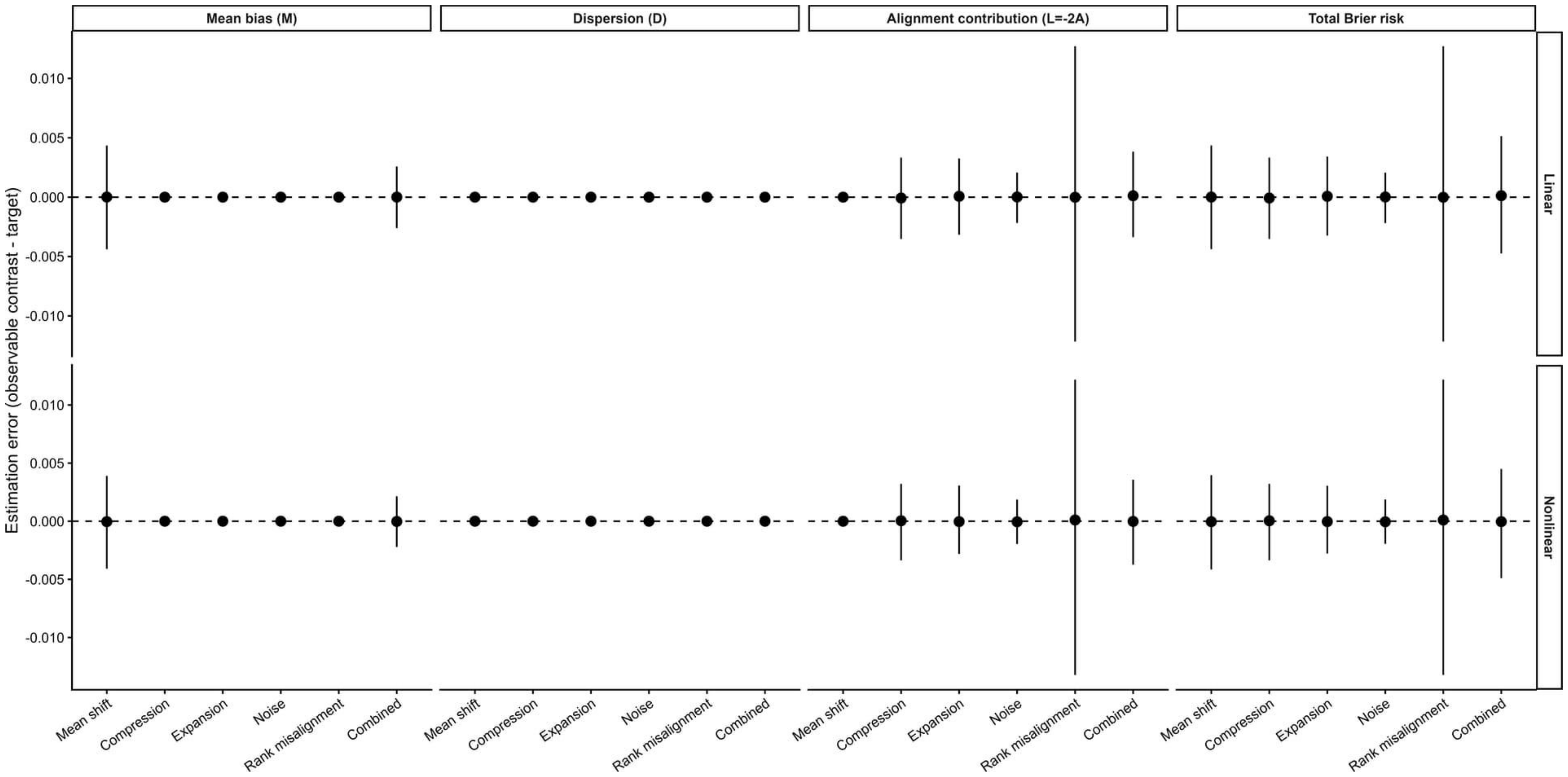
Mean bias (M)
Dispersion (D)
Alignment contribution (L=-2A)
Total Brier risk
Linear
Nonlinear
Estimation error (observable contrast - target)
0.010
0.005
0.000
-0.005
-0.010
Mean shift
Compression
Expansion
Noise
Rank misalignment
Combined

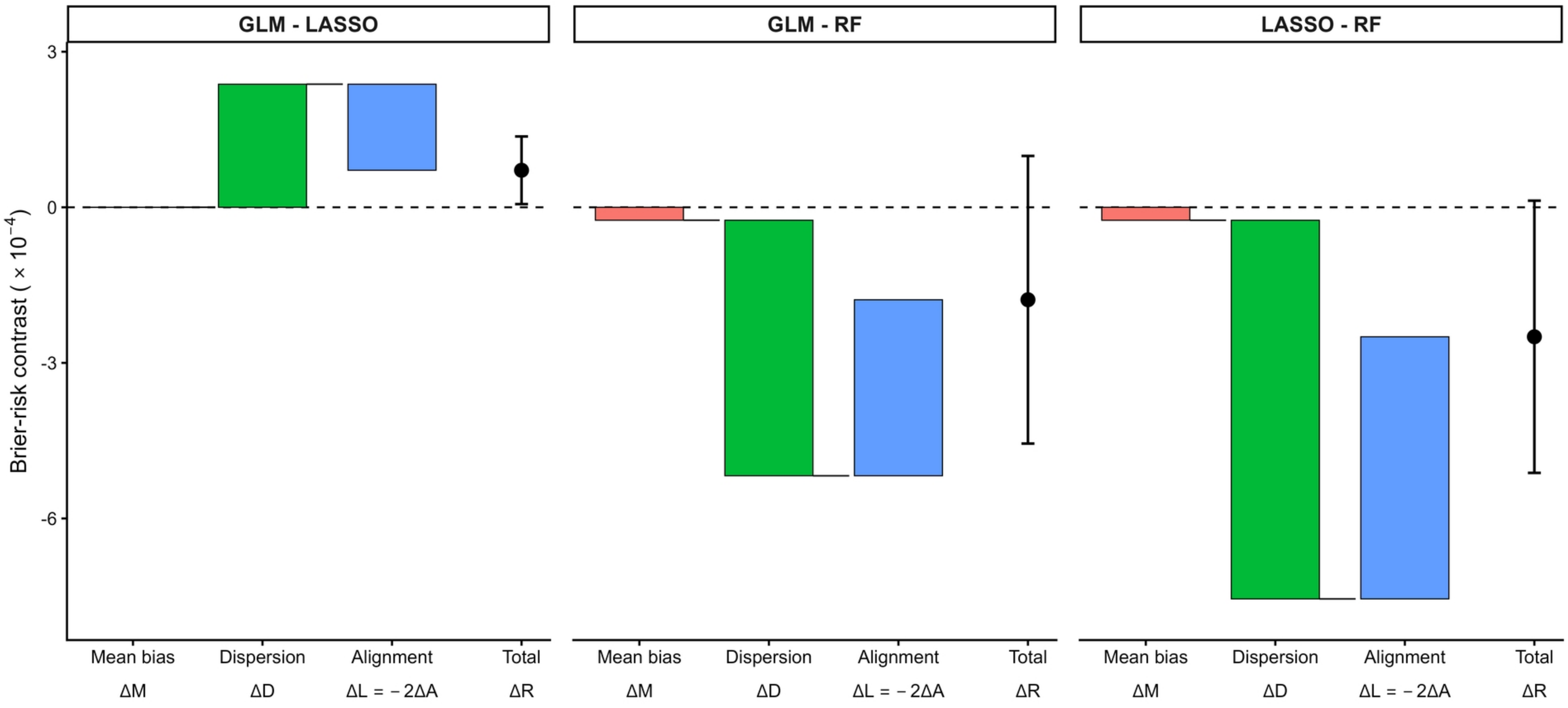
GLM - LASSO
GLM - RF
LASSO - RF
Brier-risk contrast ( × 10⁻⁴)
3
0
-3
-6
Mean bias
ΔM
Dispersion
ΔD
Alignment
ΔL = − 2ΔA
Total
ΔR

**Table 1.** Simulation design and candidate prediction rules for the component-interpretation experiment.

| Category | Item | Specification |
|---|---|---|
| General design | Sample size | $n = 1{,}500$ per replicate |
| | Monte Carlo replicates | 1,000 per setting |
| | Covariates | $X_1,\ldots,X_7$ were independently generated from $N(0,1)$. |
| Data-generating mechanism | Linear setting | $\eta = -1.0 + 0.8X_1 - 0.6X_2 + 0.5X_3 + 0.4X_4 - 0.3X_5$ |
| | Nonlinear setting | Linear predictor $+0.6(X_1^2 - 1) + 0.5X_2X_3$ |
| | True risk | $r(X) = \text{expit}(\eta)$ |
| | Outcome | $Y \mid X \sim \text{Bernoulli}\{r(X)\}$ |
| Prediction rule | Oracle | $p = r$ |
| | Mean shift | $p = \text{clip}(r + 0.10)$ |
| | Compression | $p = \mu + 0.60(r - \mu)$ |
| | Expansion | $p = \text{clip}\{\mu + 1.40(r - \mu)\}$ |
| | Noise | $p = \text{clip}(r + 0.05X_7)$ |
| | Rank misalignment | Reassignment of r according to the ranks of $X_6$ |
| | Combined | $p = \text{clip}\{\mu + 1.25(r - \mu) + 0.06 + 0.35(p_{\text{mis}} - \mu)\}$ |
| Evaluation | Oracle quantities | $M, D, L, R$ |
| | Observable contrasts | $\Delta M, \Delta D, \Delta L, \Delta R$ |
| | Monte Carlo summaries | Mean estimation error and 2.5th–97.5th percentiles of its replicate distribution |
| | Identity check | Component sum equals the paired empirical Brier-risk difference within each replicate |

$\mu$ denotes the empirical mean of the conditional event risks within a replicate. $N(0,1)$denotes the standard normal distribution, and $\text{expit}(z) = 1/\{1 + \exp(-z)\}$. $p_{\text{mis}}$ denotes the risk values reassigned according to the ranks of $X_6$ for the rank-misalignment rule. $M$, $D$, $L = -2A$, and $R$ denote the mean-bias contribution, dispersion contribution, Brier-scale alignment contribution, and Brier risk, respectively, where $A$denotes the alignment term. $\Delta$ denotes the contrast between the candidate rule and the oracle predictor (candidate rule minus oracle). The clipping operator constrains predicted probabilities to $[10^{-6}, 1 - 10^{-6}]$. The candidate rules were designed to create interpretable perturbation patterns and were not intended to imply that the three structural components can generally be manipulated independently. $X_6$ was used only for the rank-misalignment rule, and $X_7$ was used only for the noise rule; neither covariate was included in the outcome-generating model.

**Table 2. Oracle Brier-UDAM contributions and mean empirical Brier-risk contrasts under controlled perturbations.**

| Setting | Prediction rule | $M$ | $D$ | $L = -2A$ | Mean empirical Brier-risk contrast |
|---|---|---|---|---|---|
| Linear | Oracle | 0.00000 | 0.00000 | 0.00000 | 0.00000 |
| | Mean shift | 0.00998 | −0.00015 | 0.00015 | 0.00999 |
| | Compression | 0.00000 | −0.02950 | 0.03687 | 0.00730 |
| | Expansion | 0.00002 | 0.03658 | −0.03092 | 0.00576 |
| | Noise | 0.00000 | 0.00206 | 0.00034 | 0.00242 |
| | Rank misalignment | 0.00000 | 0.00000 | 0.09220 | 0.09219 |
| | Combined | 0.00352 | 0.02749 | −0.01994 | 0.01119 |
| Nonlinear | Oracle | 0.00000 | 0.00000 | 0.00000 | 0.00000 |
| | Mean shift | 0.00961 | −0.00247 | 0.00261 | 0.00971 |
| | Compression | 0.00000 | −0.03819 | 0.04774 | 0.00959 |
| | Expansion | 0.00000 | 0.03601 | −0.03073 | 0.00525 |
| | Noise | 0.00000 | 0.00148 | 0.00088 | 0.00232 |
| | Rank misalignment | 0.00000 | 0.00000 | 0.11935 | 0.11946 |
| | Combined | 0.00289 | 0.02637 | −0.01791 | 0.01131 |

Monte Carlo values are means over 1,000 replicates with n=1,500 per replicate. $M$, $D$, and $L = -2A$ are oracle risk-based contributions for each rule relative to the oracle predictor. The last column is the mean empirical Brier-risk difference, calculated solely from observed outcomes and paired predictions. $M$ is non-negative; $D$ and the Brier-scale alignment contribution $L = -2A$ are signed, and only $M + D + L$ is necessarily non-negative. Clipping prevents perfect component isolation. These results are from the replicate-specific component-interpretation experiment and should be distinguished from the fixed-rule inference experiment in the Supplementary Material.

**Table 3. Predictive performance of the fitted prediction rules in the held-out validation sample.**

| Model | Brier score [95% CI] | AUC [95% CI] | Calibration intercept [95% CI] | Calibration slope [95% CI] |
|---|---|---|---|---|
| GLM | 0.07991 [0.07700, 0.08282] | 0.620 [0.607, 0.632] | −0.022 [−0.068, 0.024] | 0.796 [0.710, 0.883] |
| LASSO | 0.07984 [0.07692, 0.08275] | 0.621 [0.608, 0.633] | −0.022 [−0.067, 0.024] | 0.883 [0.792, 0.975] |
| Random forest | 0.08009 [0.07722, 0.08295] | 0.618 [0.606, 0.630] | −0.066 [−0.111, −0.020] | 0.713 [0.636, 0.791] |

AUC, area under the receiver operating characteristic curve; CI, confidence interval; GLM, generalized linear model; LASSO, least absolute shrinkage and selection operator.

Performance was evaluated in the common held-out validation sample. Confidence intervals quantify sampling uncertainty conditional on the fitted prediction rules. Higher AUC indicates better discrimination, whereas lower Brier score indicates better overall probabilistic prediction performance. A calibration intercept of 0 and calibration slope of 1 indicate ideal calibration on the specified logistic recalibration scale.

**Table 4. Observable Brier-UDAM decomposition of paired Brier-risk contrasts in the held-out validation sample.**

| Comparison | Mean bias ($\Delta M$) | Dispersion ($\Delta D$) | Alignment contribution ($\Delta L = -2\Delta A$) | Brier risk ($\Delta R$) |
|---|---|---|---|---|
| GLM - LASSO | 0.0000001<br>[-0.0000003, 0.0000004] | 0.0002373<br>[0.0002209, 0.0002538] | -0.0001660<br>[-0.0002316, -0.0001005] | 0.0000714<br>[0.0000062, 0.0001366] |
| GLM - RF | -0.0000249<br>[-0.0000507, 0.0000009] | -0.0004928<br>[-0.0005960, -0.0003895] | 0.0003394<br>[0.0000487, 0.0006302] | -0.0001783<br>[-0.0004556, 0.0000991] |
| LASSO - RF | -0.0000250<br>[-0.0000509, 0.0000010] | -0.0007301<br>[-0.0008276, -0.0006327] | 0.0005055<br>[0.0002303, 0.0007806] | -0.0002497<br>[-0.0005121, 0.0000128] |

$A$ denotes the alignment term. Contrasts are defined as the first prediction rule minus the second. Positive $\Delta R$ indicates a larger Brier risk for the first rule. The signed contributions sum exactly to the paired empirical Brier-risk contrast apart from numerical rounding. Confidence intervals are influence-function-based and condition on the prediction rules fitted in the independent development sample. GLM, generalized linear model; LASSO, least absolute shrinkage and selection operator; RF, random forest.

**Appendix A. Lemma underlying the observable representation of the alignment-contribution contrast**

Let $h(X)$ be any square-integrable function of $X$. Then

$$\mathrm{Cov}\{Y, h(X)\} = \mathrm{Cov}\{r(X), h(X)\},$$

where $r(X) = E(Y|X) = P(Y = 1|X)$.

**Proof**

By the law of iterated expectations,

$$\begin{aligned} E\{Yh(X)\} &= E[E\{Yh(X)|X\}] \\ &= E[h(X)E(Y|X)] \\ &= E\{r(X)h(X)\}. \end{aligned}$$

Moreover,

$$E(Y) = E\{E(Y|X)\} = E\{r(X)\}.$$

Therefore,

$$\begin{aligned} \mathrm{Cov}\,\{Y, h(X)\} &= E\{Yh(X)\} - E(Y)E\{h(X)\} \\ &= E\{r(X)h(X)\} - E\{r(X)\}E\{h(X)\} \\ &= \mathrm{Cov}\,\{r(X), h(X)\}. \end{aligned}$$

This proves the result.

**Appendix B. Direct algebraic verification of the finite-sample contrast identity**

**Proposition B1 (Exact finite-sample paired-contrast identity).**

Consider paired predictions $p_{1i}$and $p_{2i}$ for the same $n$evaluation subjects, with observed binary outcomes $Y_i$, $i = 1,...,n$. Define all empirical variances and covariances using denominator $n$. Then

$$\widehat{\Delta R_{12}} = \widehat{\Delta M_{12}} + \widehat{\Delta D_{12}} + \widehat{\Delta L_{12}}.$$

**Proof.**

The empirical Brier-risk contrast is

$$\widehat{\Delta R} = \frac{1}{n}\sum_{i=1}^{n} [(Y_i - p_{1i})^2 - (Y_i - p_{2i})^2] .$$

Expanding the squared terms gives

$$\widehat{\Delta R} = \frac{1}{n}\sum_{i=1}^{n} \left(p_{1i}^2 - p_{2i}^2\right) - \frac{2}{n}\sum_{i=1}^{n} Y_i\,(p_{1i} - p_{2i}).$$

Let

$$\overline{Y} = \frac{1}{n}\sum_{i=1}^{n} Y_i\,, \overline{p}_j = \frac{1}{n}\sum_{i=1}^{n} p_{ji}\,, j = 1,2.$$

Using

$$\frac{1}{n}\sum_{i=1}^{n} p_{ji}^2 = \mathrm{Var}_n\left(p_j\right) + \overline{p}_j^2,$$

we obtain

$$\frac{1}{n}\sum_{i=1}^{n} \left(p_{1i}^2 - p_{2i}^2\right) = \mathrm{Var}_n(p_1) - \mathrm{Var}_n(p_2) + \overline{p}_1^2 - \overline{p}_2^2.$$

Similarly,

$$\frac{1}{n}\sum_{i=1}^{n} Y_i\,(p_{1i} - p_{2i}) = \mathrm{Cov}_n\{Y, p_1 - p_2\} + \overline{Y}\left(\overline{p}_1 - \overline{p}_2\right).$$

Therefore,

$$\begin{aligned}\widehat{\Delta R} &= \mathrm{Var}_n(p_1) - \mathrm{Var}_n(p_2) \\ &\quad + \overline{p}_1^2 - \overline{p}_2^2 - 2\overline{Y}\left(\overline{p}_1 - \overline{p}_2\right) \\ &\quad -2\mathrm{Cov}_n\{Y, p_1 - p_2\}.\end{aligned}$$

Because

$$\left(\bar{p}_1 - \bar{Y}\right)^2 - \left(\bar{p}_2 - \bar{Y}\right)^2 = \bar{p}_1^2 - \bar{p}_2^2 - 2\bar{Y}\left(\bar{p}_1 - \bar{p}_2\right),$$

the preceding expression becomes

$$\widehat{\Delta R} = \left[\left(\bar{p}_1 - \bar{Y}\right)^2 - \left(\bar{p}_2 - \bar{Y}\right)^2\right] + \left[\mathrm{Var}_n(p_1) - \mathrm{Var}_n(p_2)\right] - 2\mathrm{Cov}_n\{Y, p_1 - p_2\}.$$

Recognizing the three terms as the empirical mean-bias, dispersion, and alignment contribution contrasts,

$$\widehat{\Delta M} = \left(\bar{p}_1 - \bar{Y}\right)^2 - \left(\bar{p}_2 - \bar{Y}\right)^2,$$

$$\widehat{\Delta D} = \mathrm{Var}_n(p_1) - \mathrm{Var}_n(p_2),$$

and

$$\widehat{\Delta L} = -2\mathrm{Cov}_n\{Y, p_1 - p_2\},$$

we obtain

$$\widehat{\Delta R} = \widehat{\Delta M} + \widehat{\Delta D} + \widehat{\Delta L}.$$

This identity holds exactly in every finite sample, apart from numerical rounding, provided that all quantities are computed using the same evaluation subjects, the same paired predictions, and the same denominator convention for empirical variances and covariances. It is an algebraic identity and does not rely on asymptotic approximation.

**Appendix C. Structural evolution under nested information sets**

Let $Y \in \{0,1\}$,

$$r(X) = P(Y = 1|X),$$

and let

$$\mathcal{F}_0 \subseteq \mathcal{F}_1 \subseteq \cdots \subseteq \mathcal{F}_K$$

be nested information sets. Define

$$p_k = E(Y|\mathcal{F}_k) = E\{r(X)|\mathcal{F}_k\}.$$

Then

$$M_k = 0,$$
$$D_k = \mathrm{Var}(p_k) - \mathrm{Var}\{r(X)\},$$
$$L_k = 2[\mathrm{Var}\{r(X)\} - \mathrm{Var}\,(p_k)],$$

and therefore

$$L_k = -2D_k$$

and

$$T_k = \mathrm{Var}\{r(X)\} - \mathrm{Var}(p_k)\,.$$

Furthermore, for $k \geq 1$,

$$\Delta D_k = \mathrm{Var}(p_k) - \mathrm{Var}(p_{k-1}) \geq 0,$$
$$\Delta L_k = -2\Delta D_k,$$

and

$$\Delta T_k = -\Delta D_k.$$

**Proof**

By iterated expectation,

$$E(p_k) = E\{r(X)\},$$

so that

$$M_k = [E\{p_k - r(X)\}]^2 = 0.$$

Since $p_k$is $\mathcal{F}_k$-measurable,

$$\begin{aligned} E\{r(X)p_k\} &= E[E\{r(X)p_k|\mathcal{F}_k\}] \\ &= E[p_k E\{r(X)|\mathcal{F}_k\}] \\ &= E(p_k^2). \end{aligned}$$

Together with $E(p_k) = E\{r(X)\}$, this implies

$$\mathrm{Cov}\{r(X), p_k\} = \mathrm{Var}(p_k)\,.$$

Therefore,

$$
\begin{aligned}
L_k &= -2\mathrm{Cov}\left\{r(X), p_k - r(X)\right\} \\
&= -2\left[\mathrm{Cov}\left\{r(X), p_k\right\} - \mathrm{Var}\left\{r(X)\right\}\right] \\
&= 2\left[\mathrm{Var}\left\{r(X)\right\} - \mathrm{Var}\,(p_k)\right] \\
&= -2D_k.
\end{aligned}
$$

It follows that

$$
\begin{aligned}
T_k &= M_k + D_k + L_k \\
&= \mathrm{Var}\left\{r(X)\right\} - \mathrm{Var}\,(p_k).
\end{aligned}
$$

For nested information sets,

$$E(p_k|\mathcal{F}_{k-1}) = p_{k-1}.$$

The law of total variance therefore gives

$$\mathrm{Var}(p_k) = \mathrm{Var}(p_{k-1}) + E\left[\mathrm{Var}\,(p_k|\mathcal{F}_{k-1})\right].$$

Hence,

$$\Delta D_k = E\left[\mathrm{Var}\,(p_k|\mathcal{F}_{k-1})\right] \geq 0.$$

The identities

$$\Delta L_k = -2\Delta D_k$$

and

$$\Delta T_k = -\Delta D_k$$

follow immediately.

**Appendix D. Component changes along a linear prediction path**

Let $p_0(X)$ and $p_1(X)$ be square-integrable prediction functions, define

$$d(X) = p_1(X) - p_0(X),$$

and let

$$p_w(X) = p_0(X) + wd(X), \qquad 0 \leq w \leq 1.$$

Then

$$M(w) - M(0) = 2wE\{p_0(X) - r(X)\}E\{d(X)\} + w^2[E\{d(X)\}]^2, D(w) - D(0) = 2w \operatorname{Cov}\{p_0(X), d(X)\} + w^2 \operatorname{Var}\{d(X)\},$$

and

$$L(w) - L(0) = -2w \operatorname{Cov}\{r(X), d(X)\}.$$

Moreover,

$$T(w) = T(0) + 2wC + w^2 Q,$$

where

$$C = E[\{p_0(X) - r(X)\}d(X)]$$

and

$$Q = E\{d(X)^2\}.$$

If $Q > 0$, the minimizer over $0 \leq w \leq 1$ is

$$w^* = \Pi_{[0,1]}\left(-\frac{C}{Q}\right).$$

At an interior optimum,

$$-L'(w^*) = M'(w^*) + D'(w^*).$$

If $E\{d(X)\} = 0$, then $M'(w) = 0$ and $-L'(w^*) = D'(w^*)$.

**Proof**

The expression for the mean-bias component follows from

$$E\{p_w(X) - r(X)\} = E\{p_0(X) - r(X)\} + wE\{d(X)\}.$$

Squaring this expression and subtracting $M(0)$gives

$$M(w) - M(0) = 2wE\{p_0(X) - r(X)\}E\{d(X)\} + w^2[E\{d(X)\}]^2.$$

For the dispersion component,

$$\begin{aligned} \operatorname{Var}\{p_w(X)\} &= \operatorname{Var}\{p_0(X) + wd(X)\} \\ &= \operatorname{Var}\{p_0(X)\} + 2w\operatorname{Cov}\{p_0(X), d(X)\} \\ &+ w^2 \operatorname{Var}\{d(X)\}. \end{aligned}$$

Subtracting $D(0)$ yields the stated dispersion identity. For the alignment contribution,

$$L(w) = -2\,\mathrm{Cov}\left[r(X), p_0(X) - r(X) + wd(X)\right]$$
$$= L(0) - 2w\mathrm{Cov}\left\{r(X), d(X)\right\}.$$

Finally,

$$p_w(X) - r(X) = p_0(X) - r(X) + wd(X).$$

Expanding the squared structural error and taking expectations gives

$$T(w) = T(0) + 2wC + w^2Q.$$

Because $Q \geq 0$, $T(w)$is convex in $w$. If $Q > 0$, the unconstrained minimizer is

$$-\frac{C}{Q},$$

and projection onto $[0,1]$ gives the constrained minimizer.

At an interior optimum,

$$T'(w^*) = 0.$$

Since

$$T'(w) = M'(w) + D'(w) + L'(w),$$

it follows that

$$-L'(w^*) = M'(w^*) + D'(w^*).$$

If $E\{d(X)\} = 0$, then $M(w) = M(0)$for all $w$, and therefore

$$-L'(w^*) = D'(w^*).$$

This completes the proof.

**Supplementary Figure S1.** Empirical coverage of nominal 95% Wald confidence intervals across sample sizes.

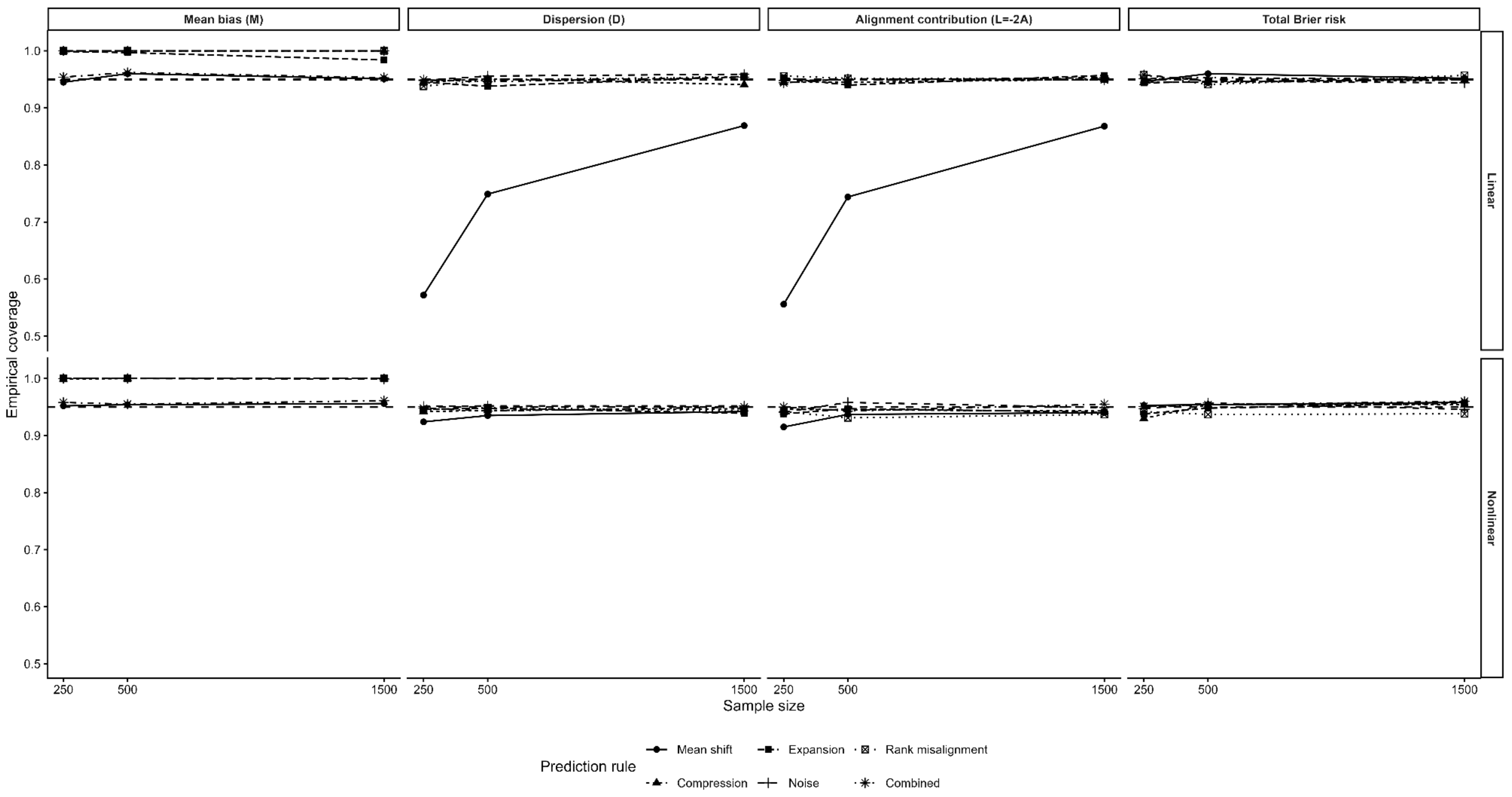

Empirical coverage is shown for the Brier-UDAM mean-bias (M), dispersion (D), Brier-scale alignment (L=-2A), and total Brier-risk contrasts under the linear and nonlinear mechanisms. Each point is based on 1,000 Monte Carlo replicates. The reference line is 0.95. Coverage for first-order-degenerate M contrasts is descriptive and should not be interpreted as evidence of valid first-order Wald inference. The marked undercoverage of the dispersion and alignment-contribution contrasts under the linear mean-shift rule arose from very small clipping-induced contrasts with poor finite-sample normal approximation; these contrasts were not algebraically first-order degenerate.

## Supplementary Table S1. Coverage and standard-error calibration for Brier-UDAM contribution contrasts.

| Setting | Prediction rule | Component | Target | First-order regularity of mean-bias contrast | n=250 | n=500 | n=1,500 |
|---|---|---|---|---|---|---|---|
| Linear | Mean shift | M | 0.009976 | Yes | 0.945 (1.013) | 0.960 (1.010) | 0.951 (1.018) |
| Linear | Mean shift | D | -0.000146 | — | 0.572 (0.687) | 0.749 (0.827) | 0.869 (0.962) |
| Linear | Mean shift | L | 0.000151 | — | 0.556 (0.676) | 0.744 (0.833) | 0.868 (0.952) |
| Linear | Mean shift | R | 0.009981 | — | 0.946 (1.013) | 0.960 (1.010) | 0.952 (1.018) |
| Linear | Compression | M | 0.000000 | No | 1.000 (1.224) | 1.000 (1.280) | 1.000 (1.291) |
| Linear | Compression | D | -0.029468 | — | 0.945 (0.996) | 0.951 (1.005) | 0.941 (1.017) |
| Linear | Compression | L | 0.036835 | — | 0.945 (0.991) | 0.945 (0.979) | 0.950 (1.000) |
| Linear | Compression | R | 0.007367 | — | 0.945 (1.001) | 0.945 (0.985) | 0.949 (0.996) |
| Linear | Expansion | M | 0.000024 | Yes | 0.999 (1.070) | 0.997 (1.009) | 0.984 (1.017) |
| Linear | Expansion | D | 0.036560 | — | 0.945 (0.989) | 0.938 (1.010) | 0.955 (1.002) |
| Linear | Expansion | L | -0.030901 | — | 0.953 (1.005) | 0.940 (0.976) | 0.957 (1.004) |
| Linear | Expansion | R | 0.005683 | — | 0.944 (1.002) | 0.946 (0.980) | 0.950 (1.002) |
| Linear | Noise | M | 0.000000 | Yes | 1.000 (1.220) | 1.000 (1.264) | 1.000 (1.208) |
| Linear | Noise | D | 0.002031 | — | 0.948 (0.972) | 0.956 (1.013) | 0.959 (1.005) |
| Linear | Noise | L | 0.000377 | — | 0.948 (1.015) | 0.949 (0.999) | 0.953 (1.010) |
| Linear | Noise | R | 0.002409 | — | 0.958 (1.030) | 0.947 (1.005) | 0.944 (1.013) |
| Linear | Rank misalignment | M | 0.000000 | No | 1.000 (1.154) | 1.000 (1.303) | 1.000 (1.263) |
| Linear | Rank misalignment | D | 0.000031 | — | 0.938 (0.968) | 0.949 (0.998) | 0.955 (1.018) |
| Linear | Rank misalignment | L | 0.092091 | — | 0.956 (1.028) | 0.952 (0.986) | 0.951 (1.015) |
| Linear | Rank misalignment | R | 0.092122 | — | 0.958 (1.031) | 0.941 (0.985) | 0.957 (1.016) |
| Linear | Combined | M | 0.003519 | Yes | 0.954 (1.011) | 0.962 (1.011) | 0.953 (1.013) |
| Linear | Combined | D | 0.027502 | — | 0.949 (1.021) | 0.946 (0.989) | 0.954 (1.017) |
| Linear | Combined | L | -0.019949 | — | 0.944 (1.009) | 0.951 (1.003) | 0.949 (0.977) |
| Linear | Combined | R | 0.011072 | — | 0.951 (1.003) | 0.953 (1.007) | 0.950 (0.980) |
| Nonlinear | Mean shift | M | 0.009609 | Yes | 0.952 (0.999) | 0.954 (1.009) | 0.956 (1.060) |
| Nonlinear | Mean shift | D | -0.002461 | — | 0.924 (1.001) | 0.935 (1.001) | 0.942 (1.025) |
| Nonlinear | Mean shift | L | 0.002599 | — | 0.915 (0.987) | 0.937 (0.998) | 0.940 (1.020) |
| Nonlinear | Mean shift | R | 0.009746 | — | 0.953 (0.999) | 0.954 (1.007) | 0.958 (1.058) |
| Nonlinear | Compression | M | 0.000000 | No | 0.999 (1.294) | 1.000 (1.256) | 1.000 (1.285) |
| Nonlinear | Compression | D | -0.038099 | — | 0.942 (1.008) | 0.944 (0.989) | 0.947 (1.008) |
| Nonlinear | Compression | L | 0.047623 | — | 0.946 (0.994) | 0.945 (0.982) | 0.943 (1.000) |
| Nonlinear | Compression | R | 0.009525 | — | 0.930 (0.978) | 0.954 (0.993) | 0.954 (1.019) |
| Nonlinear | Expansion | M | 0.000000 | Yes | 1.000 (1.286) | 1.000 (1.224) | 1.000 (1.285) |
| Nonlinear | Expansion | D | 0.035968 | — | 0.946 (1.007) | 0.948 (1.007) | 0.939 (0.978) |
| Nonlinear | Expansion | L | -0.030693 | — | 0.938 (0.989) | 0.947 (0.994) | 0.941 (0.974) |
| Nonlinear | Expansion | R | 0.005275 | — | 0.938 (0.960) | 0.948 (1.004) | 0.957 (1.034) |
| Nonlinear | Noise | M | 0.000000 | Yes | 1.000 (1.312) | 1.000 (1.173) | 0.999 (1.186) |
| Nonlinear | Noise | D | 0.001447 | — | 0.951 (1.002) | 0.952 (1.029) | 0.952 (1.027) |
| Nonlinear | Noise | L | 0.000916 | — | 0.941 (0.966) | 0.958 (1.000) | 0.949 (1.005) |
| Nonlinear | Noise | R | 0.002364 | — | 0.947 (0.991) | 0.957 (1.010) | 0.946 (0.998) |
| Nonlinear | Rank misalignment | M | 0.000000 | No | 1.000 (1.200) | 1.000 (1.194) | 1.000 (1.262) |
| Nonlinear | Rank misalignment | D | 0.000015 | — | 0.945 (0.989) | 0.949 (1.008) | 0.943 (0.991) |
| Nonlinear | Rank misalignment | L | 0.118954 | — | 0.942 (0.980) | 0.931 (0.970) | 0.937 (0.973) |
| Nonlinear | Rank misalignment | R | 0.118969 | — | 0.942 (0.999) | 0.937 (0.974) | 0.938 (0.978) |
| Nonlinear | Combined | M | 0.002890 | Yes | 0.958 (0.994) | 0.955 (1.014) | 0.961 (1.050) |
| Nonlinear | Combined | D | 0.026376 | — | 0.948 (0.999) | 0.943 (0.980) | 0.950 (1.000) |
| Nonlinear | Combined | L | -0.017912 | — | 0.950 (1.021) | 0.943 (0.981) | 0.955 (1.023) |
| Nonlinear | Combined | R | 0.011355 | — | 0.951 (1.003) | 0.954 (1.004) | 0.960 (1.055) |

Components M, D, L=-2A, and R denote the mean-bias, dispersion, Brier-scale alignment, and total Brier-risk contrasts, respectively; A denotes the alignment term. Results are based on 1,000 Monte Carlo replicates per setting. Each entry under a sample size reports empirical coverage of the nominal 95% Wald confidence interval, followed in parentheses by the ratio of the mean estimated standard error to the empirical Monte Carlo standard deviation. A ratio below 1 indicates underestimation of sampling variability. First-order regularity refers to the mean-bias contrast only. Coverage for first-order-degenerate mean-bias contrasts should not be interpreted as evidence of valid Wald inference. For D and L, the poor coverage observed for the linear mean-shift rule reflects finite-sample behavior of very small clipping-induced contrasts rather than algebraic first-order degeneracy. With 1,000 replicates, the Monte Carlo standard error of a coverage estimate near 0.95 was approximately 0.007.

**Supplementary Table S2. Complete operating characteristics for Brier-UDAM contribution contrasts.**

| **Linear setting** | | | | | | | | | | |
|---|---|---|---|---|---|---|---|---|---|---|
| **Prediction rule** | **Component** | **n** | **Target** | **Bias** | **Empirical SD** | **Mean SE** | **SE ratio** | **Coverage** | **Coverage MCSE** | **M regularity** |
| Mean shift | M | 250 | 0.009976 | 0.000066 | 0.005123 | 0.005191 | 1.013262 | 0.945 | 0.007209 | Yes |
| Mean shift | M | 500 | 0.009976 | -0.000052 | 0.003644 | 0.00368 | 1.010129 | 0.96 | 0.006197 | Yes |
| Mean shift | M | 1500 | 0.009976 | 0.000092 | 0.002084 | 0.002122 | 1.018122 | 0.951 | 0.006826 | Yes |
| Mean shift | D | 250 | -0.000146 | 0 | 0.000169 | 0.000116 | 0.686974 | 0.572 | 0.015647 | — |
| Mean shift | D | 500 | -0.000146 | 0.000001 | 0.00012 | 0.000099 | 0.827083 | 0.749 | 0.013711 | — |
| Mean shift | D | 1500 | -0.000146 | 0.000005 | 0.000067 | 0.000065 | 0.96238 | 0.869 | 0.01067 | — |
| Mean shift | L | 250 | 0.000151 | -0.000002 | 0.000187 | 0.000127 | 0.675918 | 0.556 | 0.015712 | — |
| Mean shift | L | 500 | 0.000151 | -0.000001 | 0.000131 | 0.000109 | 0.832831 | 0.744 | 0.013801 | — |
| Mean shift | L | 1500 | 0.000151 | -0.000005 | 0.000075 | 0.000071 | 0.952177 | 0.868 | 0.010704 | — |
| Mean shift | R | 250 | 0.009981 | 0.000064 | 0.005128 | 0.005195 | 1.013177 | 0.946 | 0.007147 | — |
| Mean shift | R | 500 | 0.009981 | -0.000052 | 0.003648 | 0.003684 | 1.009755 | 0.96 | 0.006197 | — |
| Mean shift | R | 1500 | 0.009981 | 0.000092 | 0.002086 | 0.002124 | 1.018194 | 0.952 | 0.00676 | — |
| Compression | M | 250 | 0 | 0.000035 | 0.000287 | 0.000351 | 1.223818 | 1 | 0 | No |
| Compression | M | 500 | 0 | 0.000014 | 0.00014 | 0.000179 | 1.280272 | 1 | 0 | No |
| Compression | M | 1500 | 0 | 0.000003 | 0.000045 | 0.000058 | 1.290528 | 1 | 0 | No |
| Compression | D | 250 | -0.029468 | 0.000087 | 0.002412 | 0.002403 | 0.996321 | 0.945 | 0.007209 | — |
| Compression | D | 500 | -0.029468 | 0.00009 | 0.001697 | 0.001705 | 1.004953 | 0.951 | 0.006826 | — |
| Compression | D | 1500 | -0.029468 | 0.000035 | 0.00097 | 0.000987 | 1.016758 | 0.941 | 0.007451 | — |
| Compression | L | 250 | 0.036835 | -0.000055 | 0.005184 | 0.005136 | 0.990871 | 0.945 | 0.007209 | — |
| Compression | L | 500 | 0.036835 | -0.000278 | 0.003718 | 0.003641 | 0.979385 | 0.945 | 0.007209 | — |
| Compression | L | 1500 | 0.036835 | 0.000057 | 0.002107 | 0.002108 | 1.000357 | 0.95 | 0.006892 | — |
| Compression | R | 250 | 0.007367 | 0.000066 | 0.004214 | 0.004218 | 1.000912 | 0.945 | 0.007209 | — |
| Compression | R | 500 | 0.007367 | -0.000174 | 0.003039 | 0.002995 | 0.985487 | 0.945 | 0.007209 | — |
| Compression | R | 1500 | 0.007367 | 0.000095 | 0.001732 | 0.001726 | 0.995991 | 0.949 | 0.006957 | — |
| Expansion | M | 250 | 0.000024 | 0.000022 | 0.00037 | 0.000396 | 1.070342 | 0.999 | 0.000999 | Yes |
| Expansion | M | 500 | 0.000024 | 0.000009 | 0.000227 | 0.000229 | 1.008817 | 0.997 | 0.001729 | Yes |
| Expansion | M | 1500 | 0.000024 | 0.000009 | 0.000114 | 0.000115 | 1.017122 | 0.984 | 0.003968 | Yes |
| Expansion | D | 250 | 0.03656 | -0.000139 | 0.002827 | 0.002796 | 0.988745 | 0.945 | 0.007209 | — |
| Expansion | D | 500 | 0.03656 | -0.000094 | 0.001972 | 0.001992 | 1.010131 | 0.938 | 0.007626 | — |
| Expansion | D | 1500 | 0.03656 | -0.000031 | 0.001149 | 0.001151 | 1.001758 | 0.955 | 0.006556 | — |
| Expansion | L | 250 | -0.030901 | 0.00007 | 0.004528 | 0.004551 | 1.005 | 0.953 | 0.006693 | — |
| Expansion | L | 500 | -0.030901 | 0.000242 | 0.003307 | 0.003229 | 0.976401 | 0.94 | 0.00751 | — |
| Expansion | L | 1500 | -0.030901 | -0.000066 | 0.001861 | 0.001869 | 1.004323 | 0.957 | 0.006415 | — |
| Expansion | R | 250 | 0.005683 | -0.000047 | 0.003981 | 0.003987 | 1.001562 | 0.944 | 0.007271 | — |
| Expansion | R | 500 | 0.005683 | 0.000157 | 0.002897 | 0.002839 | 0.979893 | 0.946 | 0.007147 | — |
| Expansion | R | 1500 | 0.005683 | -0.000087 | 0.00163 | 0.001634 | 1.002392 | 0.95 | 0.006892 | — |
| Noise | M | 250 | 0 | 0.000016 | 0.000163 | 0.000199 | 1.219525 | 1 | 0 | Yes |
| Noise | M | 500 | 0 | 0.000007 | 0.000082 | 0.000104 | 1.263512 | 1 | 0 | Yes |
| Noise | M | 1500 | 0 | 0.000001 | 0.00003 | 0.000036 | 1.207985 | 1 | 0 | Yes |
| Noise | D | 250 | 0.002031 | -0.000017 | 0.001342 | 0.001304 | 0.972181 | 0.948 | 0.007021 | — |

| Noise | D | 500 | 0.002031 | 0.000012 | 0.000916 | 0.000928 | 1.012932 | 0.956 | 0.006486 | — |
|---|---|---|---|---|---|---|---|---|---|---|
| Noise | D | 1500 | 0.002031 | 0.000001 | 0.000535 | 0.000538 | 1.00479 | 0.959 | 0.00627 | — |
| Noise | L | 250 | 0.000377 | -0.000023 | 0.002844 | 0.002887 | 1.015081 | 0.948 | 0.007021 | — |
| Noise | L | 500 | 0.000377 | -0.000022 | 0.002049 | 0.002048 | 0.999408 | 0.949 | 0.006957 | — |
| Noise | L | 1500 | 0.000377 | -0.000042 | 0.001173 | 0.001185 | 1.010186 | 0.953 | 0.006693 | — |
| Noise | R | 250 | 0.002409 | -0.000024 | 0.002517 | 0.002592 | 1.029778 | 0.958 | 0.006343 | — |
| Noise | R | 500 | 0.002409 | -0.000003 | 0.001828 | 0.001837 | 1.005266 | 0.947 | 0.007085 | — |
| Noise | R | 1500 | 0.002409 | -0.00004 | 0.001047 | 0.00106 | 1.012881 | 0.944 | 0.007271 | — |
| Rank misalignment | M | 250 | 0 | 0.000395 | 0.00117 | 0.001351 | 1.154273 | 1 | 0 | No |
| Rank misalignment | M | 500 | 0 | 0.000173 | 0.000526 | 0.000685 | 1.302869 | 1 | 0 | No |
| Rank misalignment | M | 1500 | 0 | 0.000058 | 0.000178 | 0.000225 | 1.26253 | 1 | 0 | No |
| Rank misalignment | D | 250 | 0.000031 | 0.000054 | 0.005503 | 0.005328 | 0.968301 | 0.938 | 0.007626 | — |
| Rank misalignment | D | 500 | 0.000031 | 0.000059 | 0.003774 | 0.003768 | 0.998395 | 0.949 | 0.006957 | — |
| Rank misalignment | D | 1500 | 0.000031 | -0.000057 | 0.002142 | 0.002182 | 1.0185 | 0.955 | 0.006556 | — |
| Rank misalignment | L | 250 | 0.092091 | -0.000393 | 0.017451 | 0.017931 | 1.027509 | 0.956 | 0.006486 | — |
| Rank misalignment | L | 500 | 0.092091 | -0.000681 | 0.012889 | 0.012703 | 0.985562 | 0.952 | 0.00676 | — |
| Rank misalignment | L | 1500 | 0.092091 | 0.000045 | 0.007249 | 0.007357 | 1.01499 | 0.951 | 0.006826 | — |
| Rank misalignment | R | 250 | 0.092122 | 0.000057 | 0.016692 | 0.017204 | 1.030699 | 0.958 | 0.006343 | — |
| Rank misalignment | R | 500 | 0.092122 | -0.000449 | 0.012351 | 0.01217 | 0.985388 | 0.941 | 0.007451 | — |
| Rank misalignment | R | 1500 | 0.092122 | 0.000045 | 0.006917 | 0.007028 | 1.016177 | 0.957 | 0.006415 | — |
| Combined | M | 250 | 0.003519 | 0.00007 | 0.003132 | 0.003167 | 1.011269 | 0.954 | 0.006624 | Yes |
| Combined | M | 500 | 0.003519 | -0.000021 | 0.002216 | 0.00224 | 1.010686 | 0.962 | 0.006046 | Yes |
| Combined | M | 1500 | 0.003519 | 0.000065 | 0.001274 | 0.00129 | 1.012692 | 0.953 | 0.006693 | Yes |
| Combined | D | 250 | 0.027502 | -0.000084 | 0.002591 | 0.002645 | 1.020732 | 0.949 | 0.006957 | — |
| Combined | D | 500 | 0.027502 | -0.000045 | 0.001893 | 0.001872 | 0.988891 | 0.946 | 0.007147 | — |
| Combined | D | 1500 | 0.027502 | -0.000043 | 0.001063 | 0.001082 | 1.01744 | 0.954 | 0.006624 | — |
| Combined | L | 250 | -0.019949 | -0.000038 | 0.005009 | 0.005055 | 1.009177 | 0.944 | 0.007271 | — |
| Combined | L | 500 | -0.019949 | 0.000194 | 0.003568 | 0.003579 | 1.002926 | 0.951 | 0.006826 | — |
| Combined | L | 1500 | -0.019949 | -0.000053 | 0.002117 | 0.002069 | 0.977334 | 0.949 | 0.006957 | — |
| Combined | R | 250 | 0.011072 | -0.000053 | 0.00596 | 0.005976 | 1.002588 | 0.951 | 0.006826 | — |
| Combined | R | 500 | 0.011072 | 0.000127 | 0.004209 | 0.004241 | 1.007385 | 0.953 | 0.006693 | — |
| Combined | R | 1500 | 0.011072 | -0.000031 | 0.002496 | 0.002446 | 0.979821 | 0.95 | 0.006892 | — |

**Nonlinear setting**

| Prediction rule | Component | n | Target | Bias | Empirical SD | Mean SE | SE ratio | Coverage | Coverage MCSE | M regularity |
|---|---|---|---|---|---|---|---|---|---|---|
| Mean shift | M | 250 | 0.009609 | -0.00015 | 0.004892 | 0.004888 | 0.99912 | 0.952 | 0.00676 | Yes |
| Mean shift | M | 500 | 0.009609 | -0.000065 | 0.003426 | 0.003456 | 1.008562 | 0.954 | 0.006624 | Yes |
| Mean shift | M | 1500 | 0.009609 | -0.000066 | 0.001888 | 0.002 | 1.059515 | 0.956 | 0.006486 | Yes |
| Mean shift | D | 250 | -0.002461 | -0.000025 | 0.000884 | 0.000885 | 1.00084 | 0.924 | 0.00838 | — |
| Mean shift | D | 500 | -0.002461 | -0.000036 | 0.000635 | 0.000636 | 1.00115 | 0.935 | 0.007796 | — |
| Mean shift | D | 1500 | -0.002461 | 0.00003 | 0.000357 | 0.000366 | 1.02537 | 0.942 | 0.007392 | — |
| Mean shift | L | 250 | 0.002599 | 0.000032 | 0.000979 | 0.000966 | 0.987091 | 0.915 | 0.008819 | — |
| Mean shift | L | 500 | 0.002599 | 0.000045 | 0.000696 | 0.000695 | 0.997659 | 0.937 | 0.007683 | — |
| Mean shift | L | 1500 | 0.002599 | -0.00003 | 0.000391 | 0.000399 | 1.019921 | 0.94 | 0.00751 | — |
| Mean shift | R | 250 | 0.009746 | -0.000144 | 0.00498 | 0.004974 | 0.998892 | 0.953 | 0.006693 | — |

| | | | | | | | | | | | |
|---|---|---|---|---|---|---|---|---|---|---|---|
| Mean shift | R | 500 | 0.009746 | -0.000056 | 0.003491 | 0.003517 | 1.007429 | 0.954 | 0.006624 | — |
| Mean shift | R | 1500 | 0.009746 | -0.000067 | 0.001923 | 0.002035 | 1.058166 | 0.958 | 0.006343 | — |
| Compression | M | 250 | 0 | 0.000037 | 0.0003 | 0.000388 | 1.293947 | 0.999 | 0.000999 | No |
| Compression | M | 500 | 0 | 0.00002 | 0.000155 | 0.000195 | 1.2559 | 1 | 0 | No |
| Compression | M | 1500 | 0 | 0.000006 | 0.000049 | 0.000063 | 1.284704 | 1 | 0 | No |
| Compression | D | 250 | -0.038099 | -0.000157 | 0.003616 | 0.003644 | 1.007635 | 0.942 | 0.007392 | — |
| Compression | D | 500 | -0.038099 | -0.000074 | 0.002627 | 0.002599 | 0.989018 | 0.944 | 0.007271 | — |
| Compression | D | 1500 | -0.038099 | 0.000028 | 0.001484 | 0.001496 | 1.008418 | 0.947 | 0.007085 | — |
| Compression | L | 250 | 0.047623 | 0.000232 | 0.006188 | 0.006152 | 0.994208 | 0.946 | 0.007147 | — |
| Compression | L | 500 | 0.047623 | 0.000132 | 0.004452 | 0.00437 | 0.981607 | 0.945 | 0.007209 | — |
| Compression | L | 1500 | 0.047623 | -0.000001 | 0.002521 | 0.002522 | 1.000212 | 0.943 | 0.007332 | — |
| Compression | R | 250 | 0.009525 | 0.000112 | 0.004318 | 0.004225 | 0.978466 | 0.93 | 0.008068 | — |
| Compression | R | 500 | 0.009525 | 0.000078 | 0.003008 | 0.002986 | 0.992673 | 0.954 | 0.006624 | — |
| Compression | R | 1500 | 0.009525 | 0.000032 | 0.001699 | 0.00173 | 1.018762 | 0.954 | 0.006624 | — |
| Expansion | M | 250 | 0 | 0.000032 | 0.000227 | 0.000291 | 1.285737 | 1 | 0 | Yes |
| Expansion | M | 500 | 0 | 0.000007 | 0.000118 | 0.000145 | 1.223602 | 1 | 0 | Yes |
| Expansion | M | 1500 | 0 | 0.000003 | 0.000037 | 0.000048 | 1.285468 | 1 | 0 | Yes |
| Expansion | D | 250 | 0.035968 | 0.000096 | 0.002942 | 0.002963 | 1.007167 | 0.946 | 0.007147 | — |
| Expansion | D | 500 | 0.035968 | -0.00011 | 0.002091 | 0.002106 | 1.007199 | 0.948 | 0.007021 | — |
| Expansion | D | 1500 | 0.035968 | 0.000045 | 0.001248 | 0.001221 | 0.978465 | 0.939 | 0.007568 | — |
| Expansion | L | 250 | -0.030693 | -0.000081 | 0.004532 | 0.004481 | 0.988783 | 0.938 | 0.007626 | — |
| Expansion | L | 500 | -0.030693 | 0.000072 | 0.003191 | 0.003173 | 0.994405 | 0.947 | 0.007085 | — |
| Expansion | L | 1500 | -0.030693 | -0.000068 | 0.00189 | 0.00184 | 0.973582 | 0.941 | 0.007451 | — |
| Expansion | R | 250 | 0.005275 | 0.000047 | 0.003881 | 0.003725 | 0.959723 | 0.938 | 0.007626 | — |
| Expansion | R | 500 | 0.005275 | -0.000031 | 0.00262 | 0.002631 | 1.00406 | 0.948 | 0.007021 | — |
| Expansion | R | 1500 | 0.005275 | -0.000019 | 0.001475 | 0.001525 | 1.034048 | 0.957 | 0.006415 | — |
| Noise | M | 250 | 0 | 0.000006 | 0.000146 | 0.000192 | 1.312301 | 1 | 0 | Yes |
| Noise | M | 500 | 0 | 0.000003 | 0.000083 | 0.000098 | 1.172613 | 1 | 0 | Yes |
| Noise | M | 1500 | 0 | 0 | 0.000028 | 0.000034 | 1.186407 | 0.999 | 0.000999 | Yes |
| Noise | D | 250 | 0.001447 | 0.000089 | 0.001416 | 0.001419 | 1.001919 | 0.951 | 0.006826 | — |
| Noise | D | 500 | 0.001447 | 0.000004 | 0.000978 | 0.001006 | 1.028979 | 0.952 | 0.00676 | — |
| Noise | D | 1500 | 0.001447 | 0.000054 | 0.000569 | 0.000585 | 1.027209 | 0.952 | 0.00676 | — |
| Noise | L | 250 | 0.000916 | -0.000083 | 0.002945 | 0.002846 | 0.966315 | 0.941 | 0.007451 | — |
| Noise | L | 500 | 0.000916 | 0.000007 | 0.002017 | 0.002016 | 0.999507 | 0.958 | 0.006343 | — |
| Noise | L | 1500 | 0.000916 | -0.00012 | 0.001163 | 0.00117 | 1.005336 | 0.949 | 0.006957 | — |
| Noise | R | 250 | 0.002364 | 0.000013 | 0.002505 | 0.002484 | 0.991477 | 0.947 | 0.007085 | — |
| Noise | R | 500 | 0.002364 | 0.000013 | 0.001732 | 0.00175 | 1.010425 | 0.957 | 0.006415 | — |
| Noise | R | 1500 | 0.002364 | -0.000066 | 0.001018 | 0.001015 | 0.997507 | 0.946 | 0.007147 | — |
| Rank misalignment | M | 250 | 0 | 0.000487 | 0.001287 | 0.001545 | 1.200219 | 1 | 0 | No |
| Rank misalignment | M | 500 | 0 | 0.000254 | 0.000653 | 0.00078 | 1.193667 | 1 | 0 | No |
| Rank misalignment | M | 1500 | 0 | 0.000081 | 0.000201 | 0.000254 | 1.262458 | 1 | 0 | No |
| Rank misalignment | D | 250 | 0.000015 | -0.000638 | 0.00813 | 0.008041 | 0.989093 | 0.945 | 0.007209 | — |
| Rank misalignment | D | 500 | 0.000015 | -0.000345 | 0.005689 | 0.005736 | 1.00827 | 0.949 | 0.006957 | — |
| Rank misalignment | D | 1500 | 0.000015 | 0.000001 | 0.003337 | 0.003308 | 0.99138 | 0.943 | 0.007332 | — |
| Rank misalignment | L | 250 | 0.118954 | 0.000906 | 0.021393 | 0.020973 | 0.980362 | 0.942 | 0.007392 | — |
| Rank misalignment | L | 500 | 0.118954 | 0.000554 | 0.015343 | 0.014876 | 0.969562 | 0.931 | 0.008015 | — |
| Rank misalignment | L | 1500 | 0.118954 | -0.000178 | 0.008835 | 0.008598 | 0.973184 | 0.937 | 0.007683 | — |
| Rank misalignment | R | 250 | 0.118969 | 0.000755 | 0.019425 | 0.019413 | 0.999422 | 0.942 | 0.007392 | — |
| Rank misalignment | R | 500 | 0.118969 | 0.000463 | 0.014106 | 0.013743 | 0.974288 | 0.937 | 0.007683 | — |
| Rank misalignment | R | 1500 | 0.118969 | -0.000096 | 0.008114 | 0.00794 | 0.978482 | 0.938 | 0.007626 | — |
| Combined | M | 250 | 0.00289 | -0.000025 | 0.002794 | 0.002777 | 0.994045 | 0.958 | 0.006343 | Yes |
| Combined | M | 500 | 0.00289 | -0.000044 | 0.001916 | 0.001943 | 1.0142 | 0.955 | 0.006556 | Yes |
| Combined | M | 1500 | 0.00289 | -0.000023 | 0.001074 | 0.001128 | 1.050346 | 0.961 | 0.006122 | Yes |
| Combined | D | 250 | 0.026376 | -0.000004 | 0.002781 | 0.002778 | 0.998887 | 0.948 | 0.007021 | — |
| Combined | D | 500 | 0.026376 | -0.000098 | 0.002016 | 0.001975 | 0.979855 | 0.943 | 0.007332 | — |
| Combined | D | 1500 | 0.026376 | -0.000013 | 0.001142 | 0.001142 | 0.999628 | 0.95 | 0.006892 | — |
| Combined | L | 250 | -0.017912 | 0.000067 | 0.005166 | 0.005274 | 1.020916 | 0.95 | 0.006892 | — |
| Combined | L | 500 | -0.017912 | 0.000089 | 0.003819 | 0.003745 | 0.980714 | 0.943 | 0.007332 | — |
| Combined | L | 1500 | -0.017912 | -0.000109 | 0.002121 | 0.00217 | 1.023049 | 0.955 | 0.006556 | — |
| Combined | R | 250 | 0.011355 | 0.000037 | 0.005884 | 0.005902 | 1.00293 | 0.951 | 0.006826 | — |
| Combined | R | 500 | 0.011355 | -0.000054 | 0.004139 | 0.004158 | 1.004488 | 0.954 | 0.006624 | — |
| Combined | R | 1500 | 0.011355 | -0.000145 | 0.002283 | 0.00241 | 1.0554 | 0.96 | 0.006197 | — |

Results are based on 1,000 Monte Carlo replicates per setting. Bias is the Monte Carlo mean estimate minus the reference target. Empirical SD is the Monte Carlo standard deviation of the estimates, mean SE is the average influence-function-based standard error, and SE ratio is mean SE divided by empirical SD. Components M, D, L, and R denote mean bias, dispersion, the Brier-scale alignment contribution $L=-2A$, and the total Brier-risk contrast, respectively. First-order regularity refers to the mean-bias contrast only; an em dash indicates that this classification is not applicable. Coverage for first-order-degenerate mean-bias contrasts should not be interpreted as evidence of valid first-order Wald inference. Coverage MCSE is the Monte Carlo standard error of the estimated coverage probability.

**Supplementary Figure S2. Calibration curves and predicted-risk distributions.**

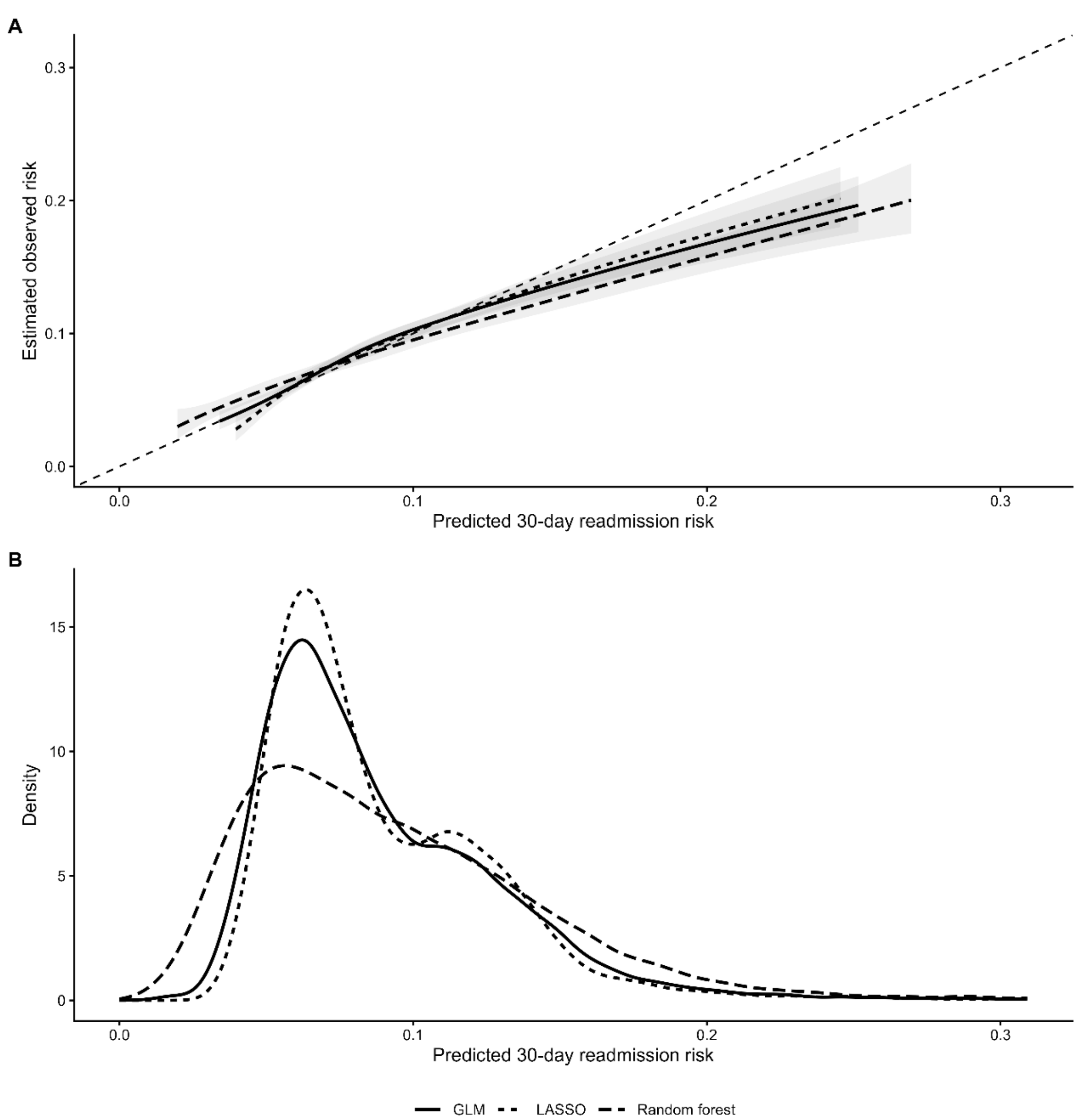

Panel A shows flexible calibration curves in the held-out validation sample; the diagonal line represents ideal calibration, and the shaded bands represent pointwise 95% confidence intervals from the fitted spline calibration models. Panel B shows the distributions of predicted 30-day readmission risks. Line types distinguish GLM, LASSO, and random forest. Abbreviations: GLM, generalized linear model; LASSO, least absolute shrinkage and selection operator.

**Supplementary Table S3. Mean-bias projection sensitivity analysis in the held-out validation sample.**

| Comparison | Mean-bias estimate (ΔM) | Wald lower | Wald upper | Projection lower | Projection upper | Gradient norm |
|---|---|---|---|---|---|---|
| GLM - LASSO | $6.77 \times 10^{-8}$ | -$2.85 \times 10^{-7}$ | $4.20 \times 10^{-7}$ | -$6.38 \times 10^{-7}$ | $1.01 \times 10^{-6}$ | 0.004946098 |
| GLM - RF | -$2.49 \times 10^{-5}$ | -$5.07 \times 10^{-5}$ | $8.90 \times 10^{-7}$ | -$5.80 \times 10^{-5}$ | $7.07 \times 10^{-6}$ | 0.013203319 |
| LASSO - RF | -$2.50 \times 10^{-5}$ | -$5.09 \times 10^{-5}$ | $9.57 \times 10^{-7}$ | -$5.82 \times 10^{-5}$ | $7.17 \times 10^{-6}$ | 0.013213846 |

The mean-bias estimate (ΔM) is the paired mean-bias contrast. Wald lower and upper and projection lower and upper are the limits of the corresponding nominal 95% confidence intervals. Gradient norm is the estimated Euclidean norm of the first-order gradient of the mean-bias contrast; smaller values indicate closer proximity to the first-order-degenerate region.

GLM, generalized linear model; LASSO, least absolute shrinkage and selection operator; RF, random forest.

**Supplementary Methods S1. Projection-based inference for a first-order-degenerate mean-bias contrast**

**Factorization and singular point**

For two fixed prediction rules evaluated on the same independent observations, define their signed mean errors by $\delta_1 = E\{p_1(X) - Y\}$ and $\delta_2 = E\{p_2(X) - Y\}$. The mean-bias contrast can be factored into two observable population means:

$$\Delta M_{12} = {\delta_1}^2 - {\delta_2}^2 = ab,$$
$$a = \delta_1 - \delta_2 = E\{p_1(X) - p_2(X)\},$$
$$b = \delta_1 + \delta_2 = E\{p_1(X) + p_2(X) - 2Y\}.$$

The gradient of $g(a,b) = ab$ is $(b,a)$. Ordinary first-order inference is therefore singular only when $a = b = 0$, equivalently $\delta_1 = \delta_2 = 0$. A null contrast with only one zero factor, such as $a = 0$ and $b \neq 0$, remains first-order regular.

**Joint confidence ellipse and projection**

Let $U_\mathrm{i} = p_{1\mathrm{i}} - p_{2\mathrm{i}}$, $V_\mathrm{i} = p_{1\mathrm{i}} + p_{2\mathrm{i}} - 2Y_\mathrm{i}$, $\theta = (a,b)^\mathrm{T}$, and $\hat{\theta} = (\bar{\mathrm{U}},\bar{V})^\mathrm{T}$. Let $\Sigma$ be the empirical 2×2 covariance matrix of $(U_\mathrm{i},V_\mathrm{i})^\mathrm{T}$. For confidence level $1 - \alpha$, define

$$C_{1-\alpha} = \{\theta : n(\hat{\theta} - \theta)^\mathrm{T}\hat{\Sigma}^{-1}(\hat{\theta} - \theta) \leq \chi^2_{2,1-\alpha}\}.$$

The projection interval for the mean-bias contrast is

$$I_{1-\alpha} = \left[\min_{(a,b)^\mathrm{T} \in C_{1-\alpha}} ab\,,\ \max_{(a,b)^\mathrm{T} \in C_{1-\alpha}} ab\right].$$

**Proposition S1**

Suppose $\sqrt{n}(\hat{\theta} - \theta)$ converges in distribution to a mean-zero bivariate normal distribution with positive-definite covariance matrix $\Sigma$, and $\hat{\Sigma}$ converges in probability to $\Sigma$. Then $\lim_{n\to\infty} inf\, P(\Delta M_{12} \in I_{1-\alpha}) \geq 1 - \alpha$.

**Proof.**

If $\theta \in C_{1-\alpha}$, then $g(\theta) = ab$ belongs to the image $g(C_{1-\alpha}) = I_{1-\alpha}$. Consequently, $P\{\Delta M_{12} \in I_{1-\alpha}\} \geq P(\theta \in C_{1-\alpha})$, and the latter probability converges to $1 - \alpha$ under the stated joint central limit theorem.

The interval is not asserted to have exact 1-α coverage because the many-to-one projection may be conservative. At $a = b = 0$ its width is $O_p\ (n^{-1})$, whereas at a first-order regular point it is $O_p(n^{-1/2})$. The interval is computed by parameterizing the ellipse boundary using a Cholesky factor of $\hat{\Sigma}$ and performing one-dimensional numerical minimization and maximization over the boundary; the interior stationary point $(0,0)$ is also checked when it lies inside the ellipse. If $\Sigma$ is singular, inference must instead be formulated in its estimable subspace.

**Relation to null-degenerate prediction contrasts**

Williamson et al.[16] used sample splitting to prevent cancellation between non-degenerate influence functions in a null variable-importance contrast. The Brier-UDAM singularity is different: when $\delta_1 = \delta_2 = 0$, each squared mean-error term itself has a zero first derivative. Directly estimating the two squared terms on separate samples therefore does not remove the singularity. The present construction instead exploits the finite-dimensional product representation of $\Delta M_{12}$.

**Scope**

The result targets paired contrasts between fixed prediction rules in a specified target population. If the target includes model fitting, variable selection, tuning, or prediction generation, uncertainty from the complete development procedure requires an outer validation or resampling scheme. This projection method addresses the mean-bias contrast only; it does not by itself resolve non-Gaussian finite-sample behavior of near-zero dispersion or alignment-contribution contrasts.

**Additional simulation**

We generated $X \sim Uniform(-1{,}1)$, $r(X) = expit(0.6X)$, and $Y|X \sim Bernoulli\ \{r(X)\}$. By symmetry $E(Y) = 0.5$. The fixed prediction rules were $p_1(X) = 0.5 + 0.60\ \{r(X) - 0.5\} + \delta_1$ and $p_2(X) = 0.5 + 1.00\{r(X) - 0.5\} + \delta_2$; all generated probabilities remained in $[0,1]$. Four settings were examined: exactly degenerate $(\delta_1,\delta_2) = (0,0)$; local to zero $(\delta_1,\delta_2) = (0.50/\sqrt{n}, 0.20/\sqrt{n})$; one-factor null $(0.02, -0.02)$, for which $b = 0$ but $a \neq 0$; and regular (0.06,0.01). For n=250, 500, and 1,500, 5,000 Monte Carlo samples were generated. We compared nominal 95% ordinary first-order Wald intervals with the 95% projection intervals.

Projection-interval coverage ranged from 0.981 to 1.000 and did not fall below the nominal level in any evaluated setting. Conservatism was greatest at the singular intersection, which is expected because $\Delta M_{12} = 0$ whenever either factor is zero. In the exactly degenerate setting, the median projection-interval width decreased approximately in proportion to $n^{-1}$. The ordinary Wald interval retained near-nominal coverage in the first-order regular settings but had coverage essentially equal to 1 at the singular point, where its usual first-order justification fails. Full operating characteristics are reported in Supplementary Table S4.

**Supplementary Table S4. Operating characteristics of projection-based inference for the mean-bias contrast.**

| Scenario | Sample size | Target ΔM | Wald coverage | Projection coverage | Wald width | Projection width |
|---|---|---|---|---|---|---|
| Exactly degenerate | 250 | 0 | 1 | 0.9998 | 0.000615 | 0.001329 |
| Exactly degenerate | 500 | 0 | 0.9996 | 0.9994 | 0.000307 | 0.000664 |
| Exactly degenerate | 1500 | 0 | 1 | 0.9998 | 0.0001 | 0.000219 |
| Local to zero | 250 | 0.00084 | 0.9568 | 0.985 | 0.004677 | 0.006032 |
| Local to zero | 500 | 0.00042 | 0.9554 | 0.9874 | 0.002338 | 0.003016 |
| Local to zero | 1500 | 0.00014 | 0.955 | 0.985 | 0.000784 | 0.001011 |
| One-factor null | 250 | 0 | 0.947 | 0.984 | 0.009777 | 0.012309 |
| One-factor null | 500 | 0 | 0.9444 | 0.985 | 0.006903 | 0.008658 |
| One-factor null | 1500 | 0 | 0.9546 | 0.9884 | 0.003992 | 0.004993 |
| Regular | 250 | 0.0035 | 0.9512 | 0.9812 | 0.012235 | 0.015362 |
| Regular | 500 | 0.0035 | 0.9478 | 0.9848 | 0.008663 | 0.010848 |
| Regular | 1500 | 0.0035 | 0.9446 | 0.9818 | 0.005003 | 0.006254 |

Coverage is the proportion of 5,000 Monte Carlo intervals containing the population target. The Monte Carlo standard error of a coverage estimate near 0.95 is approximately 0.003. The projection interval is obtained by mapping the 95% joint Wald ellipse for (a,b) through ΔM=ab; under the conditions of Proposition S1, its asymptotic coverage is at least 95%, and it may therefore be conservative.